\documentclass[11pt, fullpage]{article}
\usepackage{fullpage}
\usepackage{setspace}
\usepackage{authblk}

\usepackage[colorlinks,citecolor=blue]{hyperref}

\usepackage{cite}
\usepackage{amsmath,amssymb,amsfonts}
\usepackage{algorithmic}
\usepackage{graphicx}
\usepackage{textcomp}
\usepackage[caption=false]{subfig}
\usepackage[flushleft]{threeparttable}
\def\BibTeX{{\rm B\kern-.05em{\sc i\kern-.025em b}\kern-.08em
    T\kern-.1667em\lower.7ex\hbox{E}\kern-.125emX}}    
\allowdisplaybreaks
\usepackage{multirow}
\usepackage{xcolor}
\usepackage{enumitem}

\makeatletter
\usepackage{algorithm}
\newcommand\fs@norules{\def\@fs@cfont{\mathbfseries}\let\@fs@capt\floatc@ruled
	\def\@fs@pre{}%
	\def\@fs@post{}%
	\def\@fs@mid{\kern3pt}%
	\let\@fs@iftopcapt\iftrue}
\makeatother
\floatstyle{norules}
\restylefloat{algorithm}
\allowdisplaybreaks

\begin{document}
\title{Analysis of Screening Effects on Terahertz Photoconductive Devices using a Fully-Coupled Multiphysics Approach}
\author{Liang Chen and Hakan Bagci}


\maketitle

\begin{abstract}
The terahertz current generated by a photoconductive device (PCD) saturates as the power of the input optical pump is increased. This behavior is induced by various screening effects that stem from the interactions between electromagnetic (EM) fields and semiconductor carriers. In this work, these screening effects are numerically analyzed for the first time using a fully-coupled multiphysics approach. Unlike the previously developed simulation frameworks, this approach rigorously models the nonlinear coupling between the EM fields and the carriers and therefore is capable of accounting for the screening effects. It is demonstrated that the results obtained using this multiphysics approach and actual experiments are in excellent agreement. The optical- and radiation-field screening effects are identified in the simulation results and the optical-field screening is found to play a more dominant role in the saturation of the PCD output under high optical pump power levels.
\end{abstract}


\section{Introduction}
\label{SecIntro}
Terahertz (THz) frequency electromagnetic (EM) fields have numerous applications ranging from wireless communications to imaging systems and nondestructive testing, to material characterization~\cite{Lee2018, Saeedkia2008, Lepeshov2017review, Burford2017review, Kang2018review, Yardimci2018review, Yachmenev2019review, Ramer2015, Torkaman2018, Moon2021}. Despite the increasing interest in using THz EM fields in these applications, efficient THz source generation is still a fundamental challenge that limits the development of THz technologies. Among a variety of THz sources, photocondutive devices (PCDs) are most widely used since they are compact and frequency-stable, can be operated at room temperatures, have a wide continuous bandwidth, and can be excited using a pulsed or continuous-wave optical pump~\cite{Lepeshov2017review, Burford2017review, Kang2018review, Yardimci2018review, Yachmenev2019review}. However, it is well-known that PCDs suffer from low optical-to-THz conversion efficiency~\cite{Lepeshov2017review, Burford2017review, Kang2018review, Yardimci2018review, Yachmenev2019review}. On one hand, the portion of the optical pump power that can enter the active region of the device is limited due to the high permittivity of the photoconductive material. On the other hand, as the power of the optical power is increased, the THz radiation power generated by the PCD saturates, which limits not only the efficiency but also the maximum available power output~\cite{Lepeshov2017review, Burford2017review, Kang2018review, Yardimci2018review, Yachmenev2019review}. The former issue has been alleviated by the recent development of nanostructured PCDs. Metallic~\cite{Yardimci2018review} or dielectric~\cite{Yachmenev2019review} nanostructures, which are introduced on or inside the active region of the device and support plasmon or Mie resonances, significantly enhance the optical EM fields that interact with the semiconductor carriers. Furthermore, nanostructured electrodes also reduce the effective distance that the carriers have to travel~\cite{Berry2013, Yang2014}.

The saturation of the output THz radiation under high optical pump power levels has been a bottleneck in the operation of PCDs since their invention~\cite{Darrow1992, Benicewicz1994, Siebert2004, Kim2006, Rodriguez1996, Tani1997, Loata2007, Chou2013, Burford2016, Gagnon2016, Khorshidi2016, Burford2018, Torkaman2018}. This issue is even more pronounced for nanostructured PCDs due to the local enhancement of the optical EM fields~\cite{Lepeshov2017review, Berry2013, Yang2014}. The saturation behavior has been attributed to various screening effects in the literature~\cite{Darrow1992, Benicewicz1994, Rodriguez1996, Tani1997, Siebert2004, Kim2006, Loata2007, Chou2013, Gagnon2016, Burford2016, Khorshidi2016}. Among them, ``space-charge screening'' and ``radiation-field screening'' are extensively studied~\cite{Hu1990, Darrow1992, Benicewicz1994, Rodriguez1996, Tani1997, Siebert2004, Kim2006, Loata2007, Chou2013, Khorshidi2016}. Space-charge screening (also known as Coulomb screening) refers to the dampening of the bias field due to the (static Coulomb) electric field generated by the electron-hole separation~\cite{Tani1997, Siebert2004, Kim2006, Khorshidi2016, Gagnon2016}. Radiation-field screening is a consequence of the low-frequency EM fields that are generated by the photocurrents inside the device and effectively dampen the bias field~\cite{Darrow1992, Benicewicz1994, Rodriguez1996, Tani1997, Siebert2004, Kim2006, Loata2007}. A third effect is observed when the (effective) photoconductivity, which is increased by the generation of carriers via absorption of the optical EM field energy, limits the penetration of the optical EM fields into the active region of the device~\cite{Chou2013, Burford2016}. This effect, which is termed ``optical-field screening'' in this paper, is discussed briefly in~\cite{Chou2013, Burford2016}. Since the characterization of these screening effects requires probing the EM fields and the carrier densities inside the device, which is often difficult to do during experiments, most of the methods that have been developed so far to investigate the screening effects are based on phenomenological models~\cite{Darrow1992, Benicewicz1994, Siebert2004, Kim2006, Khorshidi2016}. 

Having said that, numerical tools are often used to characterize PCDs, but to be able to account for the screening effects, these tools must model the nonlinear interactions between the EM fields and the carriers. Earlier approaches developed for simulating conventional PCDs rely on the finite difference method (FDM). These approaches use semi-analytical expressions of the generation rate and do not strictly model the two-way coupling between the EM fields and the carriers~\cite{Sirbu2005, Neshat2010, Kirawanich2008, Khabiri2012, Khiabani2013, Young2014, Moreno2014}. In recent years, the finite element method (FEM), which is more flexible and accurate than FDM, has started to become the method of choice to compute the optical EM fields, especially on nanostructured PCDs~\cite{Burford2016, Mohammad2016nanoslit, Bashirpour2017, Burford2017}. But like the previous approaches, those that rely on FEM ignore the coupling from the carriers to the EM fields and therefore they cannot account for the screening effects~\cite{Burford2016}. 

More recently, a multiphysics framework has been developed to simulate PCDs~\cite{Chen2019multiphysics, Chen2020steadystate}. This framework takes into account the two-way nonlinear coupling between the EM fields and the carriers and solves the fully-coupled systems of Poisson and stationary drift-diffusion (DD) equations and time-domain Maxwell and DD equations for steady-state and transient fields and carriers, respectively. To efficiently account for the multiple space and time characteristic scales involved in the multiphysics model, discontinuous Galerkin (DG)-based schemes are used to discretize the coupled systems of equations. The efficiency of this framework has been further increased by adopting a unit-cell model that approximates the interactions on the whole device within one period of the nanostructure via carefully-designed boundary conditions~\cite{Chen2019unitcell, Chen2020efficient}. Its increased efficiency and ability to fully model the coupling between the EM fields and the carriers make this multphysics framework a prime candidate for analyzing screening effects.

In this work, the saturation behavior and screening effects pertinent to PCDs~\cite{Darrow1992, Benicewicz1994, Rodriguez1996, Tani1997, Siebert2004, Kim2006, Loata2007, Chou2013} are numerically characterized and identified for the first time using the fully and rigorously coupled multiphysics approach briefly discussed above~\cite{Chen2019multiphysics, Chen2020steadystate, Chen2019unitcell, Chen2020efficient}. It is demonstrated that the results obtained using this multiphysics approach and actual experiments are in excellent agreement. Numerical results generated using a large number of simulations clearly show the effects of optical- and radiation-field screening. The impacts of these screening effects are compared by controlling the coupling mechanism in the multiphysics model. It is demonstrated that the saturation of the output is observed only when the photocurrent is accounted for in the Maxwell equations (i.e., when the carrier effects are coupled back onto the EM fields). The low-frequency EM fields generated by this photocurrent dampens the bias electric field (radiation-field screening). Additionally, a higher carrier density level increases the effective photoconductivity, which in return limits the penetration of the optical EM fields into the active region of the device (optical-field screening). Without the coupling of the photocurrent to the Maxwell equations, the EM field interactions do not ``see'' this increasing photoconductivity. Finally, it is found that the optical-field screening plays a more dominant role in the saturation of the PCD output at high optical pump power levels.


\section{Multiphysics Model}
\label{SecMethod}
\begin{figure}[!b]
	\centerline{\includegraphics[width=0.8\columnwidth]{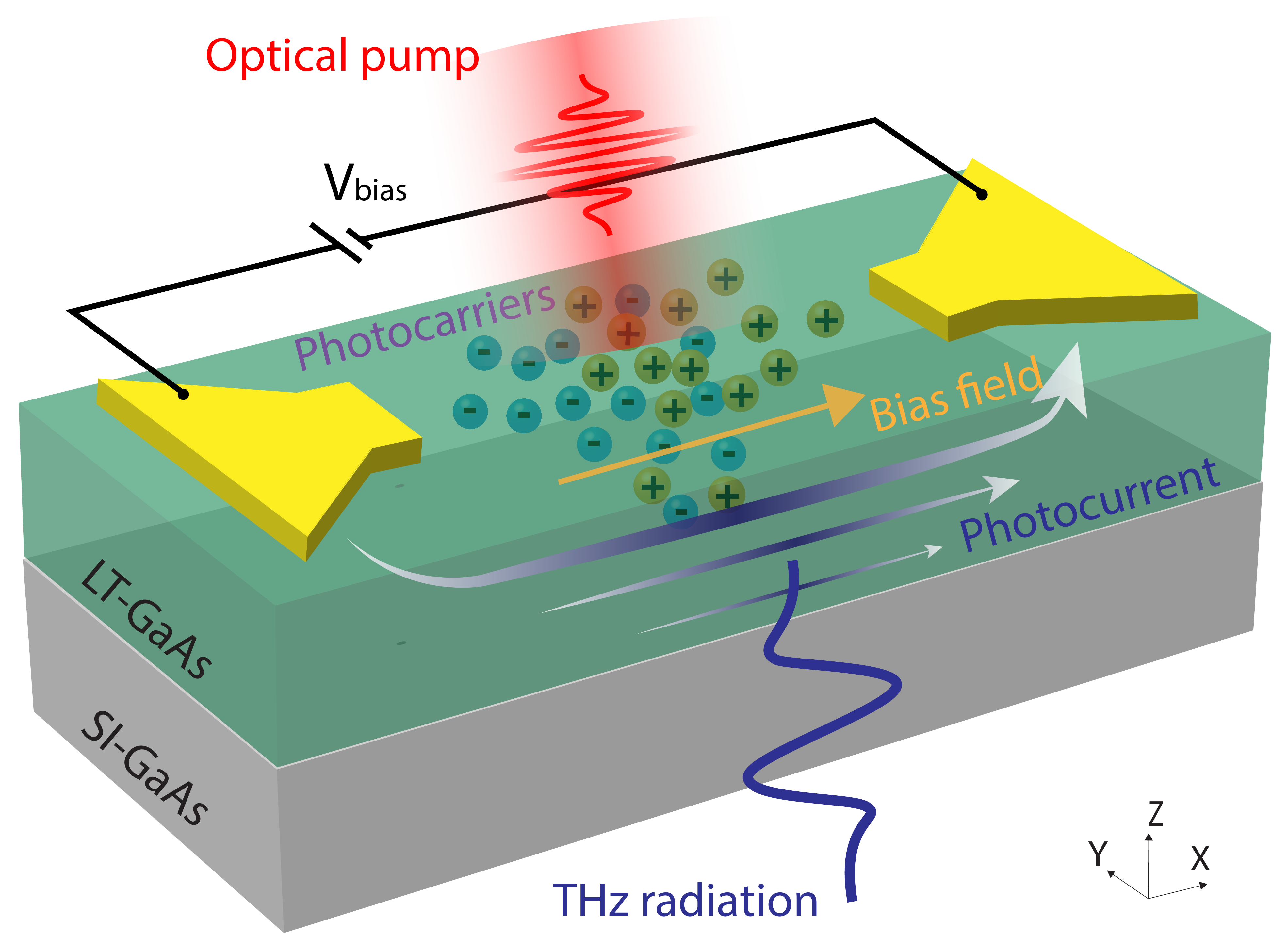}}
	\caption{Schematic illustration of the PCD operation.}
	\label{PM3Dschem}
\end{figure}

The operation of a PCD is illustrated in Fig.~\ref{PM3Dschem}. The device consists of a substrate layer (SI-GaAs), a photoconductive layer (LT-GaAs), and two electrodes that are deposited on the photoconductive layer.
The operation of a PCD has two stages. Initially, a bias voltage is applied to the electrodes. The balance between the bias electric field and the carrier distribution results in a non-equilibrium steady-state described by a coupled system of Poisson and DD equations~\cite{Vasileska2010, Chen2020steadystate}
\begin{align}
\label{s_PS} 
& \nabla  \cdot [\varepsilon ({\mathbf{r}}) {{\mathbf{E}}^s}({\mathbf{r}}) ] = q[C({\mathbf{r}}) + n_h^s({\mathbf{r}}) - n_e^s({\mathbf{r}})] \\
\label{s_DD}  
& \nabla  \cdot [ {d_c}({{\mathbf{E}}^s})\nabla n_c^s({\mathbf{r}}) \pm  {\mu _c}({{\mathbf{E}}^s}){{\mathbf{E}}^s}({\mathbf{r}})n_c^s({\mathbf{r}}) ]= {R^s}(n_e^s,n_h^s)
\end{align}
where ${{\mathbf{E}}^s({\mathbf{r}})}$ is the stationary electric field, $\varepsilon ({\mathbf{r}})$ is the dielectric permittivity, $q$ is the electron charge, $C({\mathbf{r}})$ is the doping concentration, $c\in \{ e,h\}$ represents the carrier type and hereinafter the upper and lower signs should be selected for electron ($c=e$) and hole ($c=h$), respectively, ${n_c}({\mathbf{r}},t)$ is the carrier density, ${\mu _c}({\mathbf{E}}^s)$ and ${d_c}({\mathbf{E}}^s)$ are the mobility and diffusion coefficient, respectively, and $R^s({n_e^s},{n_h^s})$ is the recombination rate~\cite{Vasileska2010, Chen2020steadystate}. The field-dependent mobility model and the recombination rate model for LT-GaAs are the same as those given in~\cite{Chen2020steadystate}.

When an optical pump laser is incident on the device, the transient stage starts. The photoconductive material absorbs the optical EM field energy and generates carriers. The carriers are driven by both the bias electric field and the optical EM fields and produce photocurrents. 
The transient interactions between the EM fields and the carriers are described by a coupled system of the time-dependent Maxwell and DD equations~\cite{Chen2019multiphysics, Chen2019discontinuous}
\begin{align}
\label{t_E} 
& \varepsilon(\mathbf{r}) {\partial _t}{{\mathbf{E}}^t}({\mathbf{r}},t) \!=\!  \nabla \times {{\mathbf{H}}^t}({\mathbf{r}},t) \!-\! [{\mathbf{J}}_e^t({\mathbf{r}},t) + {\mathbf{J}}_h^t({\mathbf{r}},t)]\\
\label{t_H} 
& \mu(\mathbf{r}) {\partial _t}{{\mathbf{H}}^t}({\mathbf{r}},t) = -\nabla \times {{\mathbf{E}}^t} ({\mathbf{r}},t) \\
\label{t_N} 
& q{\partial _t}n_c^t({\mathbf{r}},t) \!=\! \pm \nabla \cdot {\mathbf{J}}_c^t({\mathbf{r}},t) \!-\! q[{R^t}(n_e^t,n_h^t) \!-\! G({{\mathbf{E}}^t},{{\mathbf{H}}^t})]\\
\label{t_J} 
& {\mathbf{J}}_c^t({\mathbf{r}},t) = q{\mu _c}({\mathbf{E}}^s)([{{\mathbf{E}}^s}({\mathbf{r}}) + {{\mathbf{E}}^t}({\mathbf{r}},t)]n_c^t({\mathbf{r}},t) + {{\mathbf{E}}^t}({\mathbf{r}},t) n_c^s({\mathbf{r}})) \pm q{d_c}({\mathbf{E}}^s)\nabla n_c^t({\mathbf{r}},t)
\end{align}
where ${\mathbf{E}}^t({\mathbf{r}},t)$ and ${\mathbf{H}}^t({\mathbf{r}},t)$ are the time-dependent electric and magnetic fields, ${n_c^t}({\mathbf{r}},t)$ is the time-dependent carrier density, $\mu ({\mathbf{r}})$ is the permeability, ${{\mathbf{J}}_c^t}({\mathbf{r}},t)$ is the transient current densities due to carrier movement, and $R^t({n_e^t},{n_h^t})$ and $G({\mathbf{E}^t},{\mathbf{H}}^t)$ are the transient recombination and generation rates. In~\eqref{t_E}, a Lorentz dispersion model is used for $\varepsilon(\mathbf{r})$ to model the material dispersion of LT-GaAs near optical frequencies. The optical absorption is signified by the imaginary part of the permittivity, which is used in the calculation of time-dependent $G({\mathbf{E}^t},{\mathbf{H}}^t)$. Generation of the time-dependent carriers results in a time-dependent effective photoconductivity in the active region of the device.

Also, note that the photocurrent ${{\mathbf{J}}_c^t}({\mathbf{r}},t)$ in~\eqref{t_E}, which is the source of the THz EM fields, represents the coupling from the generated carriers back to the EM fields. {\color{black} The THz frequency contents present in $\mathbf{J}_{c}^{t}(\mathbf{r},t)$ stem from the term $q{{\mu }_{c}}({{\mathbf{E}}^{s}}){{\mathbf{E}}^{s}}(\mathbf{r})n_{c}^{t}(\mathbf{r},t)$ in~\eqref{t_J}: The frequency contents of $n_{c}^{t}(\mathbf{r},t)$ are in the THz frequency range, and $q$, ${{\mathbf{E}}^{s}}(\mathbf{r})$, and ${{\mu }_{c}}({{\mathbf{E}}^{s}})$ are constants in time. Since the frequency contents of the terms $q{{\mu }_{c}}({{\mathbf{E}}^{s}}){{\mathbf{E}}^{t}}(\mathbf{r},t)n_{c}^{s}(\mathbf{r})$ and $q{{\mu }_{c}}({{\mathbf{E}}^{s}}){{\mathbf{E}}^{t}}(\mathbf{r},t)n_{c}^{t}(\mathbf{r},t)$ are in the optical frequency range and the vector-component of the term $q{{d}_{c}}({{\mathbf{E}}^{s}})\nabla n_{c}^{t}(\mathbf{r},t)$, which is in the direction of the dominant component of $\mathbf{J}_{c}^{t}(\mathbf{r},t)$, is negligible, these three terms do not contribute significantly to THz frequency contents of $\mathbf{J}_{c}^{t}(\mathbf{r},t)$ and therefore do not have much effect on the THz EM field generation.}

Ignoring the source term ${{\mathbf{J}}_c^t}({\mathbf{r}},t)$ in~\eqref{t_E} (as done with the methods described in~\cite{Burford2016, Mohammad2016nanoslit, Bashirpour2017, Burford2017}) has two consequences in terms of modeling screening effects: (i) As described in Section~\ref{SecIntro} and in~\cite{Chou2013, Burford2016}, the increasing photoconductivity limits the penetration of the optical EM fields into the active region of the device, resulting in optical-field screening. If ${{\mathbf{J}}_c^t}({\mathbf{r}},t)$ in~\eqref{t_E} is ignored, the EM fields do not ``see'' the effective photoconductivity and therefore the optical-screening cannot be account for. (ii) Again, as briefly described in Section~\ref{SecIntro} and detailed in~\cite{Hu1990, Darrow1992, Benicewicz1994, Rodriguez1996, Tani1997, Siebert2004, Kim2006, Loata2007, Chou2013, Khorshidi2016}, the low-frequency EM fields dampen the bias field, resulting in radiation-field screening. Naturally, if ${{\mathbf{J}}_c^t}({\mathbf{r}},t)$ as the source of these fields is ignored in the Maxwell equations, the radition-field screening cannot be accounted for.

The Poisson-DD system~\eqref{s_PS}--\eqref{s_DD} is solved iteratively using the Gummel method~\cite{Vasileska2010} and the linearized set of equations at every iteration are discretized using a stationary discontinuous Galerkin (DG) scheme~\cite{Cockburn1998, Castillo2000, Shu2016, Chen2020steadystate, Chen2020float, Chen2020hybridizable}. The steady-state solutions are used as inputs in the transient solver~\cite{Chen2019multiphysics}. The Maxwell-DD system~\eqref{t_E}--\eqref{t_J} is discretized using a time domain DG scheme~\cite{Hesthaven2002, Lu2004, Fezoui2005, Hesthaven2008, Niegemann2009, Gedney2009, Liu2012, Sirenko2012, Chen2012discontinuous, Li2015IBC, Li2017dispersive, Sirenko2018, Liu2004, Liu2016, Harmon2016, Ren2017, Chen2019multiphysics}. The nonlinear coupling between the Maxwell equations and the DD equations is accounted for by feeding these systems' solutions to each other at alternating time steps (with different step sizes) during the explicit time marching~\cite{Chen2019multiphysics}. Multiple space and time characteristic scales are involved in the operation of a PCD, i.e., the Debye length is $\sim 10\,\mathrm{nm}$, the optical wavelength is $\sim 100\,\mathrm{nm}$, and the device size is $\sim 10\,\mathrm{\mu m}$, the optical wave period is $\sim 1\,\mathrm{fs}$ while the device response time is $\sim 1\,\mathrm{ps}$. Using a higher-order DG framework and an explicit time marching scheme that uses different time step sizes for the Maxwell equations and the DD equations helps to keep the computational cost under control~\cite{Chen2019multiphysics, Chen2020efficient}. Having said that, to further increase the efficiency of this DG-based multiphysics framework, a unit-cell model, which approximates the interactions on the whole device within one period of the nanostructure via carefully-selected/designed boundary conditions, is adopted~\cite{Chen2020efficient}. This until cell model makes it possible to run a large number of simulations within a reasonable time without sacrificing from the accuracy of the simulation results~\cite{Chen2019unitcell, Chen2020efficient}.

\section{Simulation Setup}
\label{SecSetup}
To analyze the screening effects, a conventional PCD is considered for the sake of simplifying the physical problem. The thickness of the LT-GaAs and the SI-GaAs layers is $0.5\,\mathrm{\mu m}$ and the interface between these two layers is located at $z=0$. The distance between the electrodes along the $x$ direction is $10\,\mathrm{ \mu m}$. A bias voltage $V_{\mathrm{bias}}$ is applied to the electrodes. The semiconductor parameters are provided in Table~\ref{parameters}. The permittivity of the LT-GaAs layer is expressed using the Lorentz dispersion model as
\begin{align*}
\varepsilon(\omega) = \varepsilon_{0} (\varepsilon_{\infty} + {\frac{\omega_{p}^2}{ \omega_{o}^2 - \omega^2 - i \gamma \omega } } )
\end{align*}
where $\epsilon_{\infty}=5.785$, $\omega_o=4.783\times10^{15}\,\mathrm{rad/s}$, and $\gamma=4.557\times10^{14}\,\mathrm{rad/s}$, $\omega_p=1.061\times10^{16}\,\mathrm{rad/s}$~\cite{Chen2021generation}. The electrodes are modeled as gold and their permittivity is expressed using the Drude model as
\begin{align*}
\varepsilon(\omega) = \varepsilon_{0} ( \varepsilon_{\infty} - {\frac{\omega_{p}^2}{ \omega^2 + i \gamma \omega } } )
\end{align*}
where $\epsilon_{\infty}=1.0$, $\omega_p=1.372\times10^{16}\,\mathrm{rad/s}$, and $\gamma=8.052\times10^{13}\,\mathrm{rad/s}$~\cite{Chen2019multiphysics}. SI-GaAs is modeled as a dielectric material with relative permittivity $\epsilon_r =13.26$. All materials are considered nonmagnetic $\mu_r =1.0$.

\begin{table}[!t]
	\begin{threeparttable}
		\renewcommand{\arraystretch}{1.5}
		\centering
		\caption{Semiconductor material parameters}
		\label{parameters}
		\begin{tabular}{c c}
			\hline
			C & $1.3\times 10^{16}$ cm$^{-3}$ \\
			$n_i$ & $9\times 10^6$ cm$^{-3}$ \\ 
			\hline
			Mobility &
			{ \begin{math} \begin{array}{cc}
				\mu_e^0=8000\,\mathrm{ {cm}^2/V/s }, \mu_h^0=400\,\mathrm{ {cm}^2/V/s }\\
				V_e^\mathrm{sat}\!=\!1.725 \! \times \! 10^{7}\,\mathrm{cm/s}, V_h^\mathrm{sat} \! = \! 0.9 \! \times \! 10^{7}\,\mathrm{cm/s} \\
				\beta_e=1.82, \beta_h=1.75
				\end{array} \end{math} } \\ \hline
			Recombination & 
			{ \begin{math} \begin{array}{cc}
				\tau_e=0.3\,\mathrm{ps}, \tau_h=0.4\,\mathrm{ps} \\
				n_{e1}=n_{h1}=4.5 \times 10^6\,\mathrm{cm}^{-3} \\
				C_e^A=C_h^A=7\times 10^{-30}\,\mathrm{cm}^6\mathrm{/s}
				\end{array} \end{math} } \\
			\hline
		\end{tabular}
	\end{threeparttable}
\end{table}

The DD equations are solved only within the LT-GaAs layer, while the Poisson equation and the Maxwell equations are solved everywhere. For the Poisson equation, a potential-drop boundary condition is used along the $x$ direction to mimic the bias voltage, periodic boundary conditions (PBCs) are used along the $y$ direction, and a homogeneous Neumann boundary condition is used along the $z$ direction. For the stationary DD equations, PBCs are used along $x$ and $y$ directions, and a homogeneous Robin boundary condition is used on the surfaces of the LT-GaAs layer (transverse to the $z$ direction)~\cite{Chen2020float, Chen2020hybridizable}. PBCs are used along the $x$ and $y$ directions for the time-dependent Maxwell and DD equations. Along the $z$ direction, perfectly matched layers are used for Maxwell equations~\cite{Chen2020APS_PML, Chen2020pml}, and a homogeneous Robin boundary condition is used for the DD equations~\cite{Chen2020efficient}. The simulation domains are discretized with tetrahedrons. The minimum and the maximum edge lengths in the mesh are $10\,\mathrm{nm}$ and $200\,\mathrm{nm}$, respectively.

The time-dependent electron density ($n_e^t$) and the $x$ components of the time-dependent electric field and current density ($E_x^t$ and $J_x^t$) are recorded at probes P1, P2, P3, P4, P5, P6 that are located at $x=0$, $y=0$, $z=\{10, 100, 200, 300, 400, 490\}\,\mathrm{nm}$. The Fourier transform of $E_x^t$ and $J_x^t$ is denoted by $\mathcal{F}(E_x^t)$ and $\mathcal{F}(J_x^t)$, respectively. Two types of simulations are carried out: (i) ``Coupled’’ simulation that uses the mathematical model~\eqref{t_E}--\eqref{t_J} with photocurrents $\mathbf{J}_e^t(\mathbf{r},t)$ and $\mathbf{J}_h^t(\mathbf{r},t)$ in place. (ii) ``Uncoupled’’ simulation that uses the same mathematical model except $\mathbf{J}_e^t(\mathbf{r},t)$ and $\mathbf{J}_h^t(\mathbf{r},t)$ are ignored. The uncoupled simulation is similar to the frequency-domain FEM-based approach~\cite{Burford2016, Mohammad2016nanoslit, Bashirpour2017, Burford2017} where the EM fields are computed by the Maxwell solver without taking into account the photocurrents (the screening effects are ignored). Since in the unit-cell model, the source aperture is assumed infinitely large, in the following, the peak power flux density $S_\mathrm{pump}$ (in unit of $\mathrm{mW/cm^2}$) is used as a measure of the optical pump power. Note that, due to PBCs, electrons and holes moving out of one boundary enter the unit cell from the opposite boundary. This means that the space-charge screening effect cannot be demonstrated using this model. To be able to characterize this effect, the full-device simulation~\cite{Chen2019multiphysics} should be used.

\section{Screening Effects}

{\color{black} In all simulations considered in this section, the PCD is operated in the continuous-wave mode~\cite{Lepeshov2017review, Yardimci2018review} and excited from top by two continuous-wave lasers at the same time. The operating frequencies of these lasers are $374.5~\mathrm{THz}$ and $375.5~\mathrm{THz}$, with a frequency difference of $1~\mathrm{THz}$ and their individual power flux densities are the same.}

\begin{figure}[!b]
	\centering
	\subfloat[\label{Fig5Nea}]{\includegraphics[width=0.8\columnwidth]{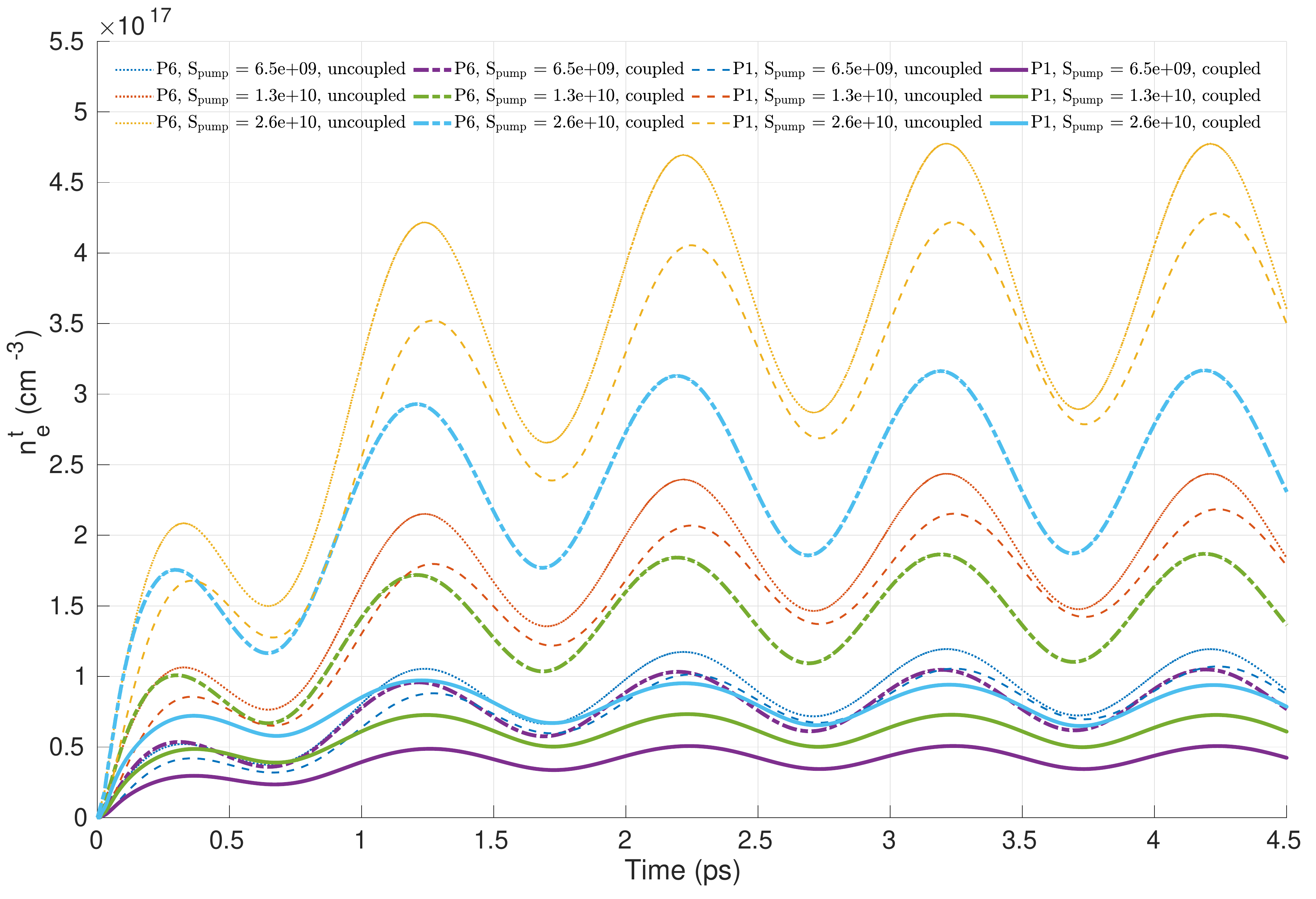}}\\
	\vspace{-0.2cm}
	\subfloat[\label{Fig5Neb}]{\includegraphics[width=0.8\columnwidth]{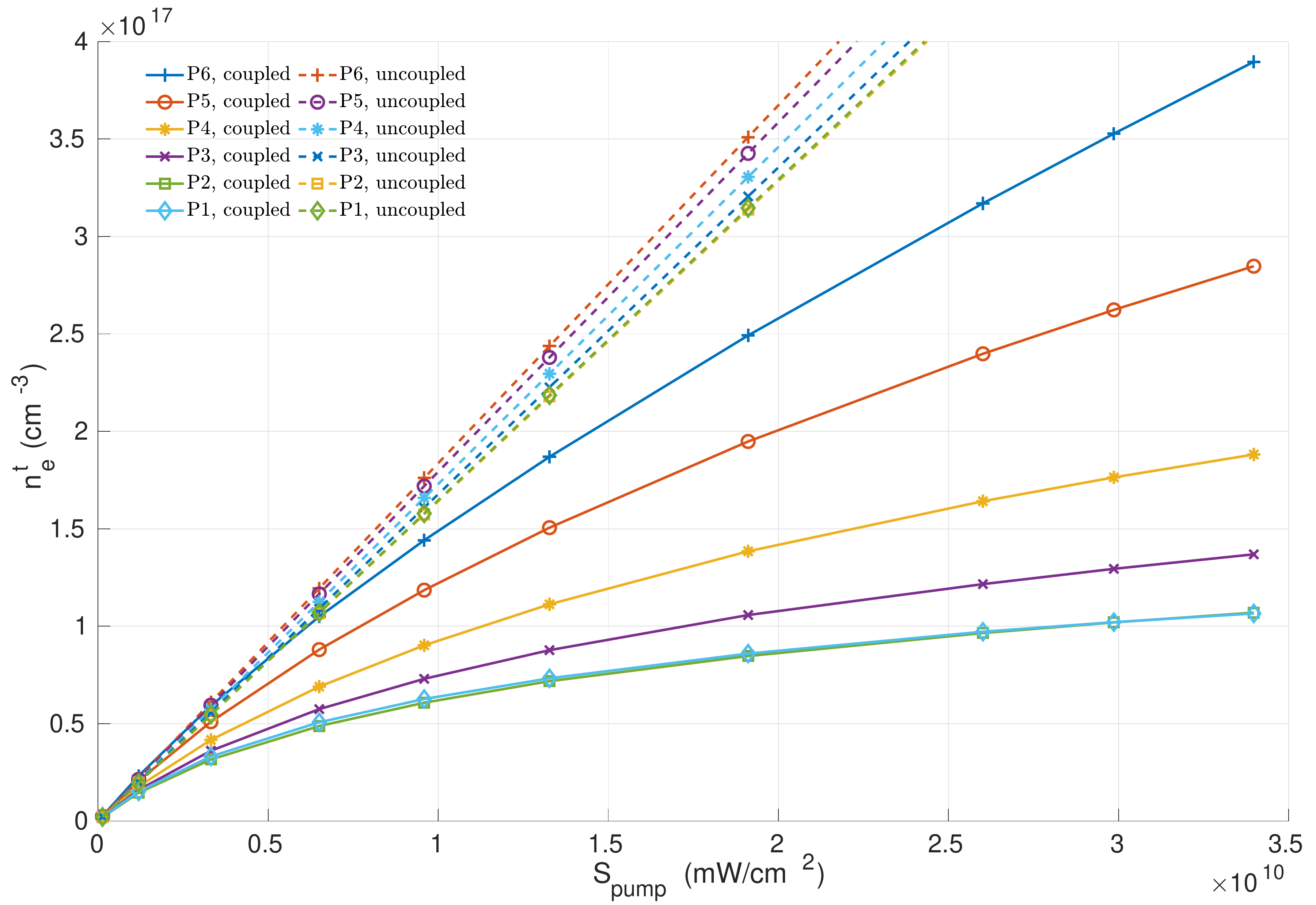}}\\
	\caption{(a) $n_e^t$ recorded at probes P6 and P1 for different values of $S_\mathrm{pump}$. (b) Maximum value of $n_e^t$ versus $S_\mathrm{pump}$. {\color{black}$S_{\mathrm{pump}}$ is in units of $\mathrm{mW/cm^2}$.}}
	\label{Fig5Ne}
\end{figure}

Fig.~\ref{Fig5Ne} (a) plots $n_e^t$ recorded at probes P6 and P1 versus time for different values of $S_\mathrm{pump}$. Fig.~\ref{Fig5Ne} (b) plots the maximum value of $n_e^t$ recorded at different probes versus $S_\mathrm{pump}$. Comparing the results obtained by the coupled and uncoupled simulations, one can clearly see that the saturation behavior results from the inclusion of the photocurrents $\mathbf{J}^t_h(\mathbf{r},t)$ and $\mathbf{J}^t_e(\mathbf{r},t)$ on the right hand side of the Maxwell equation in~\eqref{t_E}. In the coupled simulation, the increase in the maximum value of $n_e^t$ recorded at P6 (near the top, see Fig.~\ref{PM3Dschem}) slows down for higher values of $S_\mathrm{pump}$. $n_e^t$ recorded at P1 (near the bottom) is much smaller and also saturates much faster than $n_e^t$ recorded at P6. In the uncoupled simulation, the maximum value of $n_e^t$ at all probes increase linearly with increasing $S_\mathrm{pump}$.

\begin{figure}[!t]
	\centerline{\includegraphics[width=0.8\columnwidth]{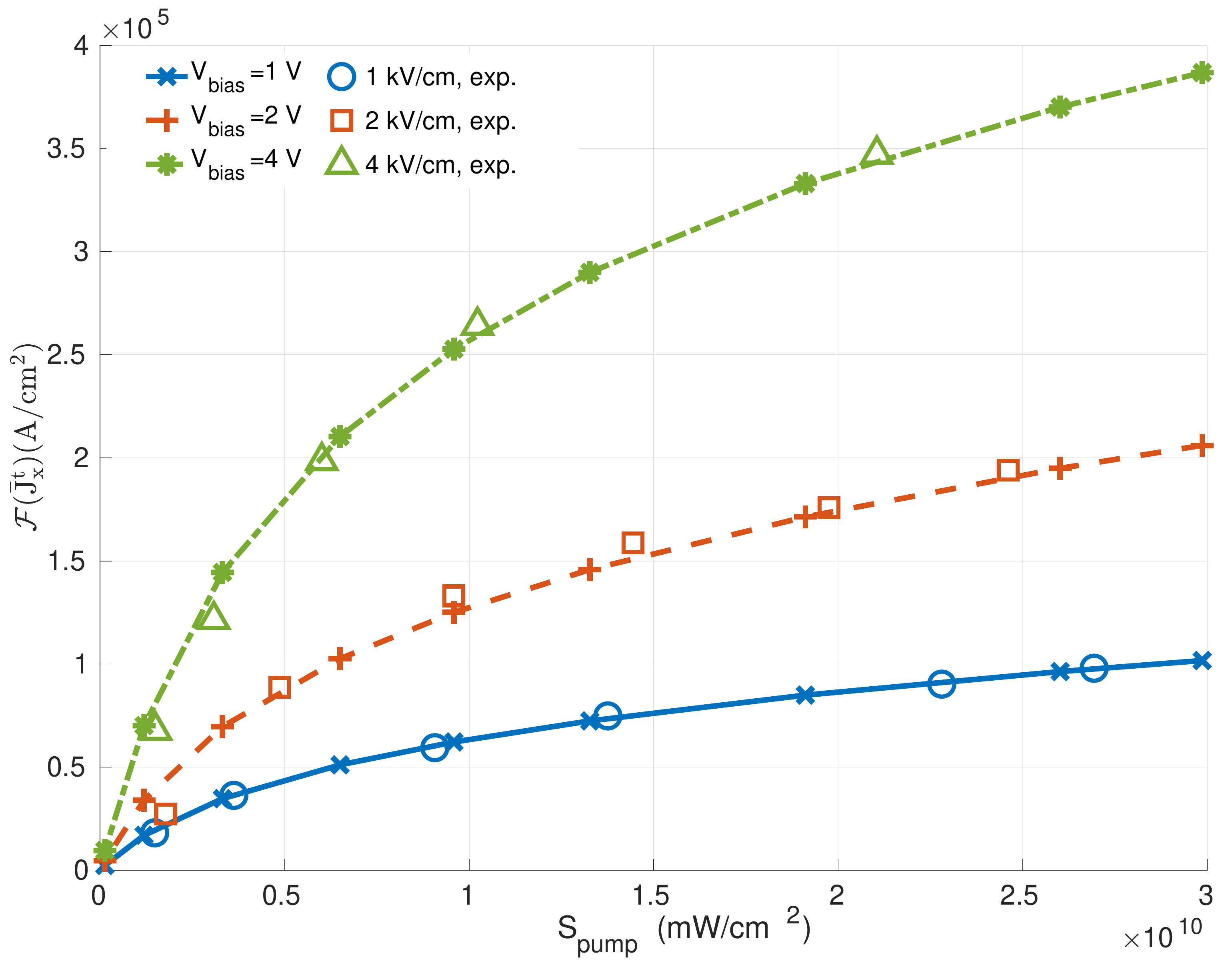}}
	\caption{$\mathcal{F}(\bar{J}_x^t)$ at $1\,\mathrm{THz}$ for different levels of optical pump power $S_\mathrm{pump}$ and different values of bias voltage $V_{\mathrm{bias}}$.}
	\label{Fig6Jx}
\end{figure}

First, the results obtained by the coupled simulation are validated against experimental data. The calibrated field data reported in~\cite{Darrow1992} is obtained from the radiated field measured during the experiments. Note that, in the Fourier domain, the magnitude of the radiated field is proportional to the magnitude of the photocurrent density~\cite{Darrow1992, Benicewicz1994, Chou2013}. Therefore, the magnitude of the calibrated field data in~\cite{Darrow1992} is expected to be proportional to the magnitude of the photocurrent density. Fig.~\ref{Fig6Jx} plots  $|\mathcal{F}(\bar{J}_x^t)|$
computed at $1\,\mathrm{THz}$ versus $S_\mathrm{pump}$ for different values of $V_\mathrm{bias}$ and compare it to the normalized calibrated field data (denoted as ``exp.'' in the figure). Here, $\bar{J}_x^t = (J_x^t|_{P4} + J_x^t|_{P5} + J_x^t|_{P6})/3$ represents the average current density recorded at probes P4, P5, and P6. Only the photocurrents at these three probes are considered for comparison because only the carriers near the top interface ($\sim100\,\mathrm{nm}$) contribute to the generation of THz EM fields (due to the short carrier lifetime)~\cite{Yardimci2018review, Yang2014}. The bias voltage and the laser power used in the simulation are calculated from the experimental parameters~\cite{Darrow1992}. Note that the normalization factor for the field data is $1.36\times 10^5$ for all values of $S_\mathrm{pump}$ and $V_\mathrm{bias}$.

Fig.~\ref{Fig6Jx} shows the excellent agreement between the simulation results and the experimental data. Noticeably, the photocurrent density does not show clear saturation with increasing $V_{\mathrm{bias}}$. This also agrees with the experimental results in~\cite{Benicewicz1994, Liu2003, Gobel2011}. Note that that similar and consistent results are obtained for $|\mathcal{F}(\bar{J}_x^t)|$ computed at frequencies within band $[0.1, 1.5] \,\mathrm{THz}$. These comparison results and observations validate the accuracy of the coupled multiphysics model used in this work.

\begin{figure}[!t]
	\centering
	\subfloat[\label{Fig7HNea}]{\includegraphics[width=0.6\columnwidth]{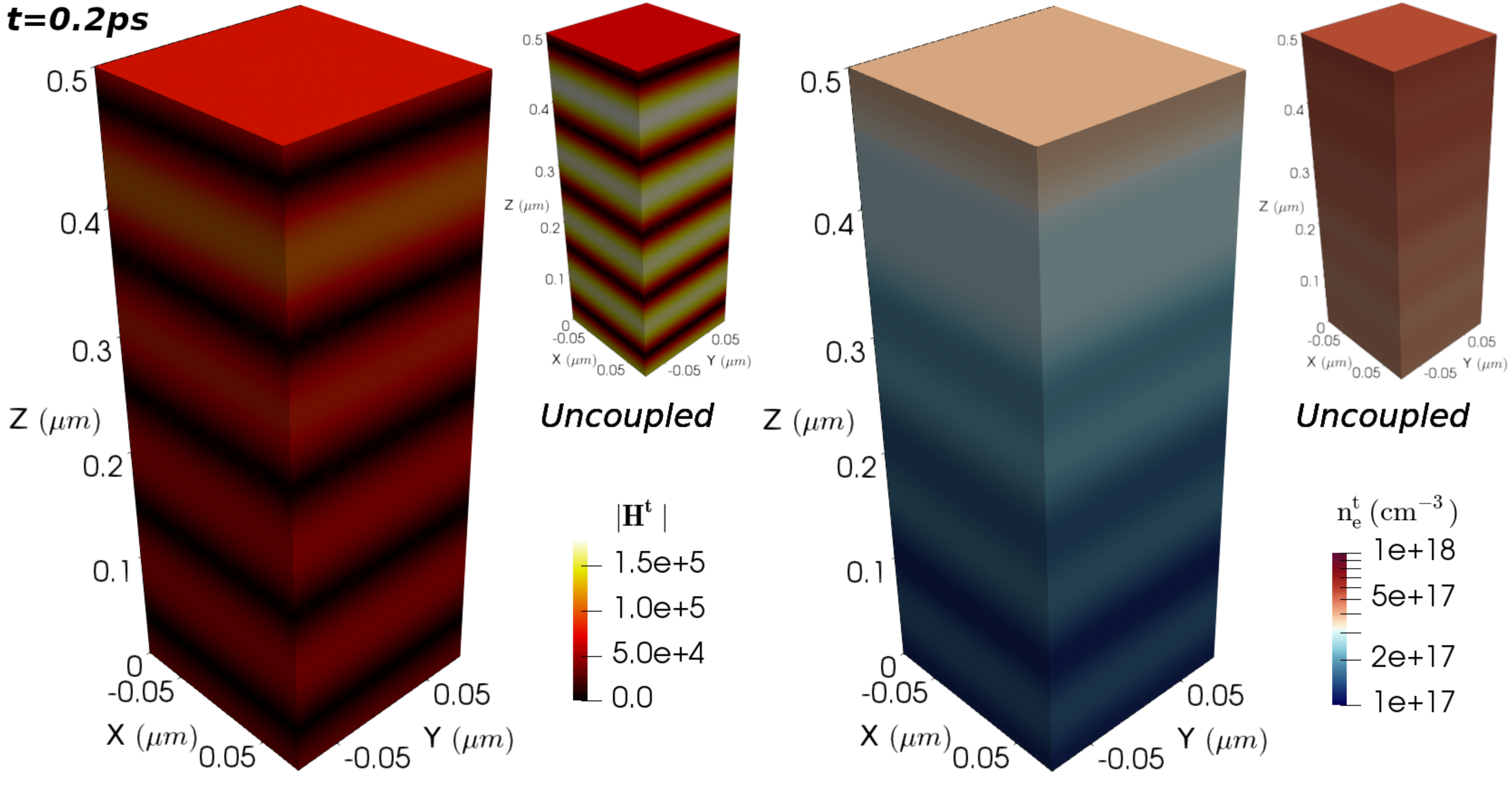}}\\
	\subfloat[\label{Fig7HNeb}]{\includegraphics[width=0.6\columnwidth]{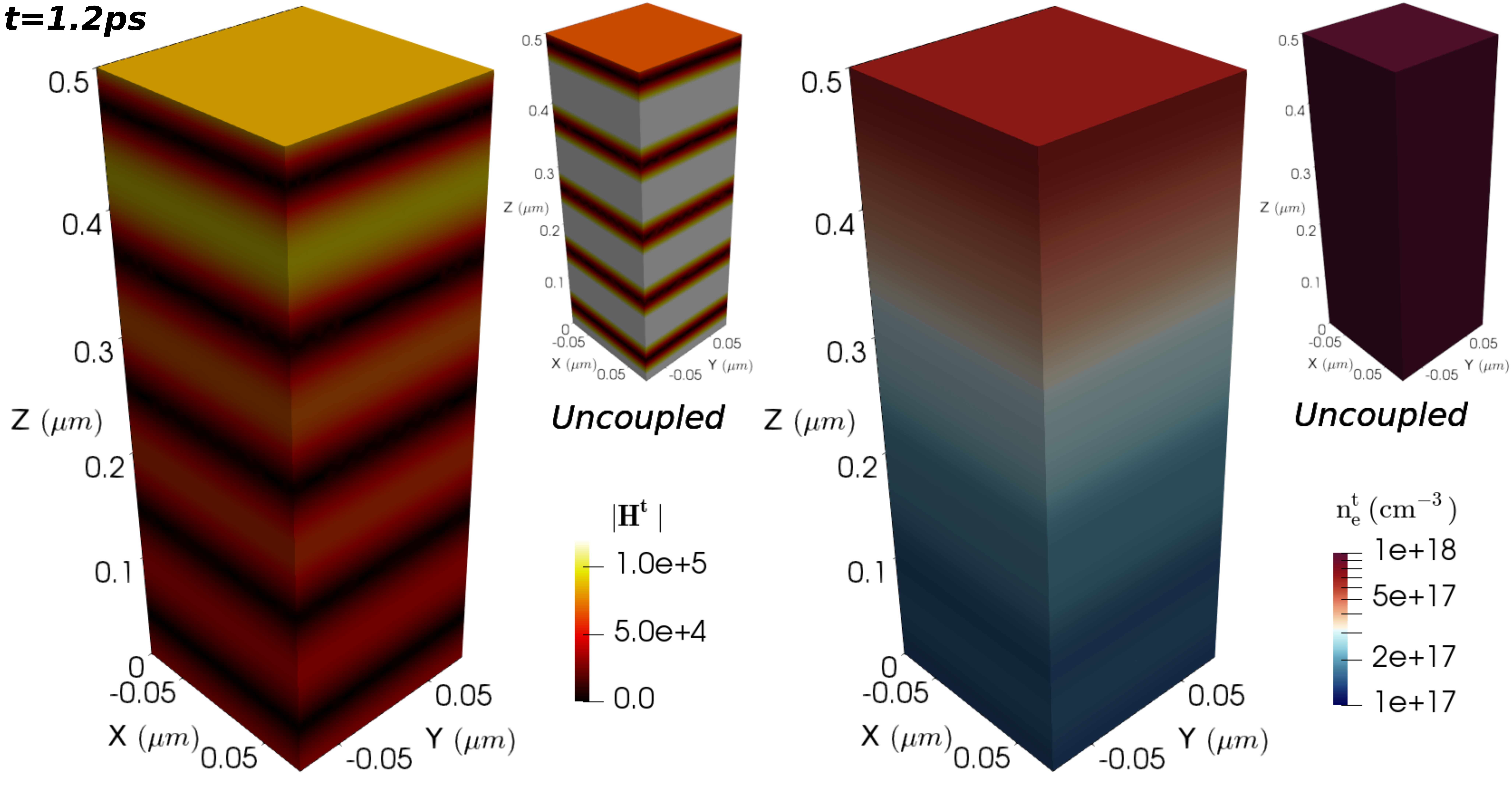}}\\
	\subfloat[\label{Fig7HNec}]{\includegraphics[width=0.6\columnwidth]{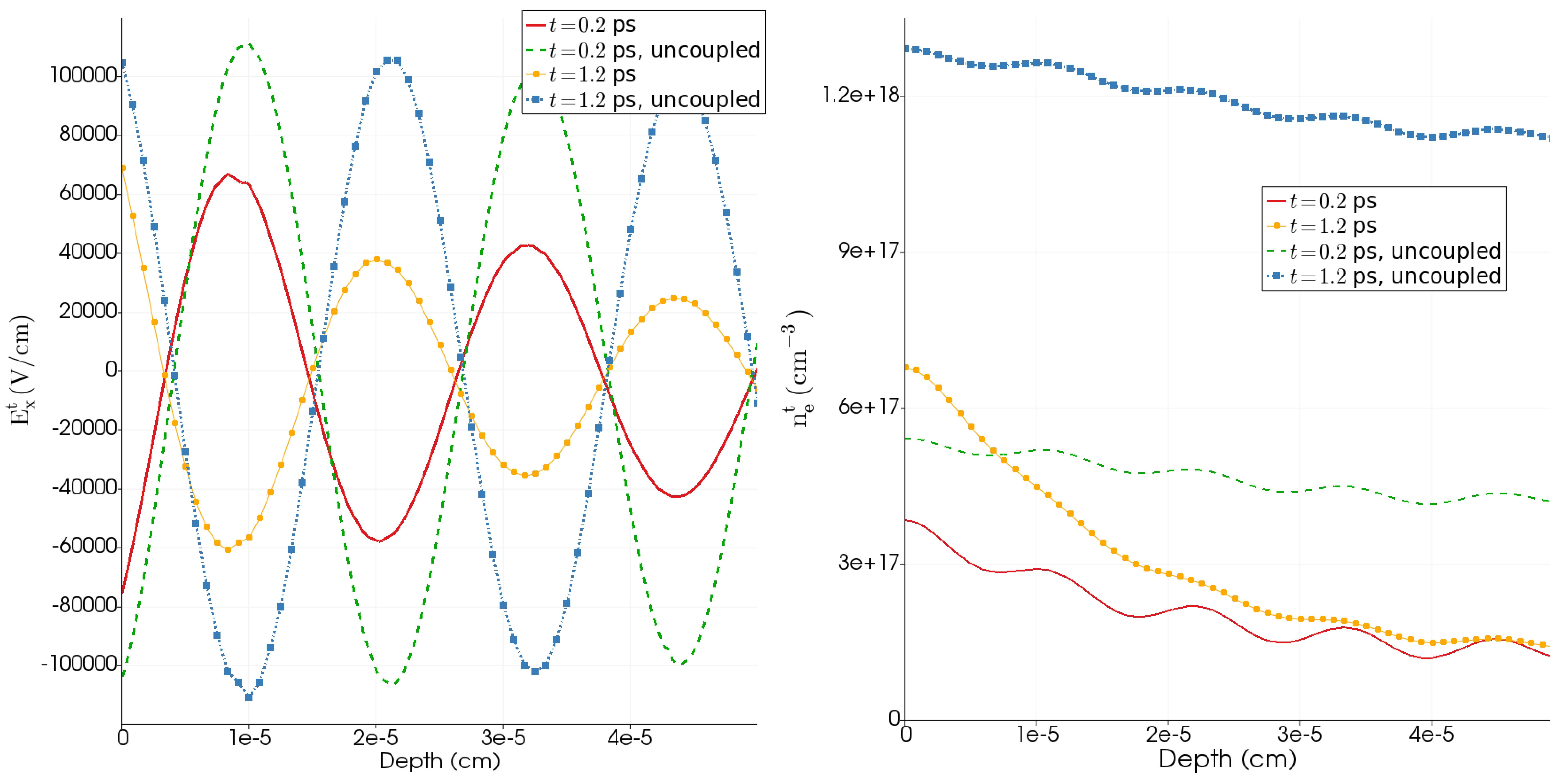}}
	\caption{Distributions of $|\mathbf{H}^t|$ and $n_e^t$ at (a) $0.2\,\mathrm{ps}$ and (b) $1.2\,\mathrm{ps}$ as computed by the coupled simulation. The insets correspond to the same results computed by the uncoupled simulation. (c) $|\mathbf{H}^t|$ and $n_e^t$ along the line $(x=0,~y=0,~0\le z \le 0.5\,\mathrm{\mu m})$.}
	\vspace{-0.1cm}
	\label{Fig7HNe}
\end{figure}

Fig.~\ref{Fig7HNe} (a) and (b) show distributions of the magnetic field's magnitude $|\mathbf{H}^t|$ and $n_e^t$ for $S_\mathrm{pump}=8.3\times 10^{10} \, \mathrm{mW/cm^2}$ at $0.2 \, \mathrm{ps}$ and $1.2 \, \mathrm{ps}$, respectively. The distribution of $|\mathbf{H}^t|$ shows a standing wave pattern along the $z$ direction due to multiple reflections in the layered structure. Since the generation rate depends on the EM field, the distribution of $n_e^t$ shows a similar standing wave pattern. 
Fig.~\ref{Fig7HNe} (c) shows $E_x^t$ and $n_e^t$ along the line $(x=0, y=0, 0\le z\le 0.5 \, \mathrm{\mu m})$. Both quantities decrease with depth (along the $-z$ direction). In the uncoupled simulation, the slow decay is due to the imaginary part of the permittivity near optical frequencies. This imaginary part represents the optical absorption that results in carrier generation~\cite{Chen2021generation}. In the results obtained by the coupled simulation, the decay is much faster and $E_x^t$ oscillates around a negative value (decaying towards it). Meanwhile, at $t=1.2 \, \mathrm{ps}$, $n_e^t$ concentrates on the top interface ($z=0.5 \, \mathrm{\mu m}$) and the standing wave pattern is smoothened.

\begin{figure}
	\centering
	\vspace{-0.1cm}
	\subfloat[\label{Fig8Exta}]{\includegraphics[width=0.5\columnwidth]{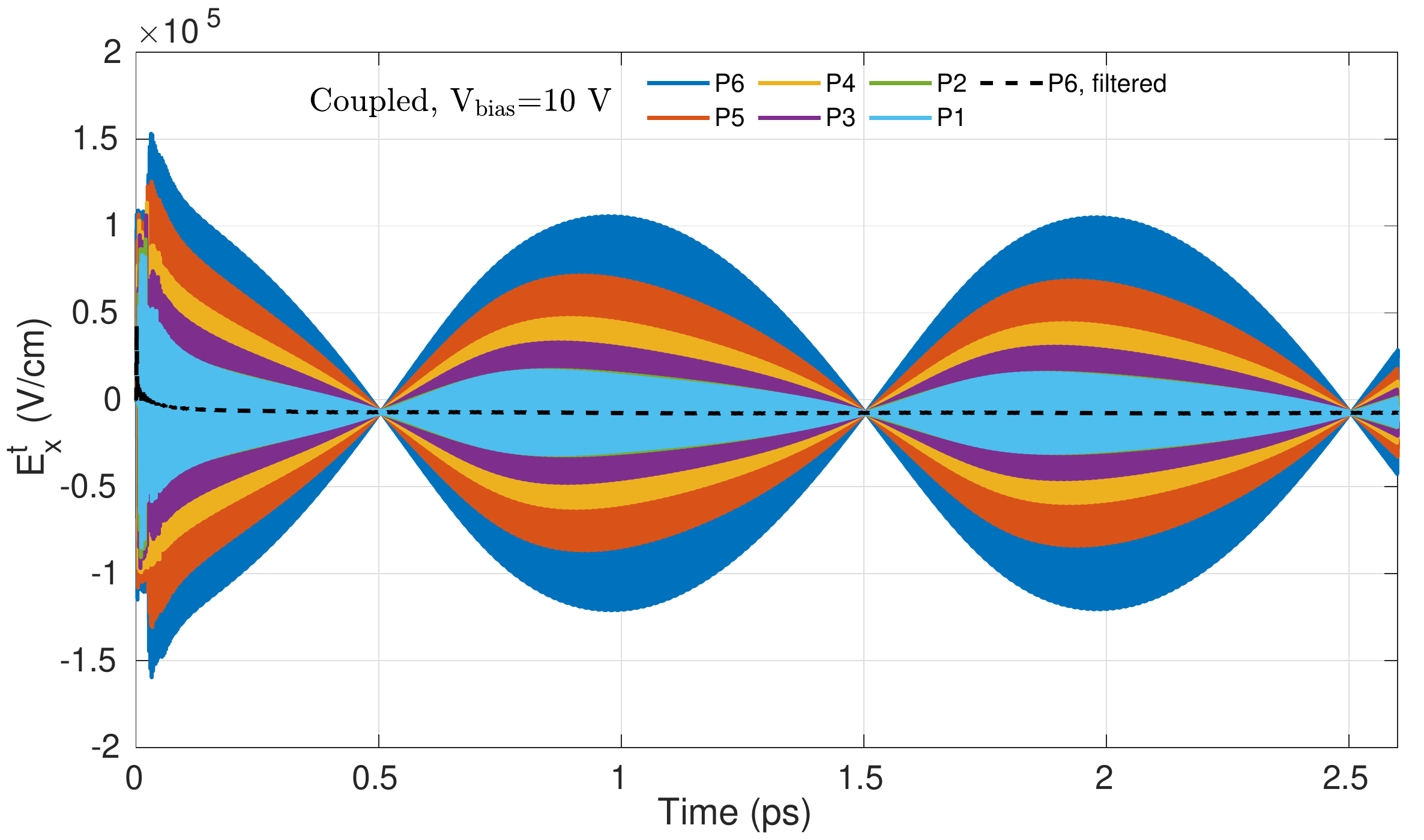}}\\
	\vspace{-0.28cm}
	\subfloat[\label{Fig8Extb}]{\includegraphics[width=0.5\columnwidth]{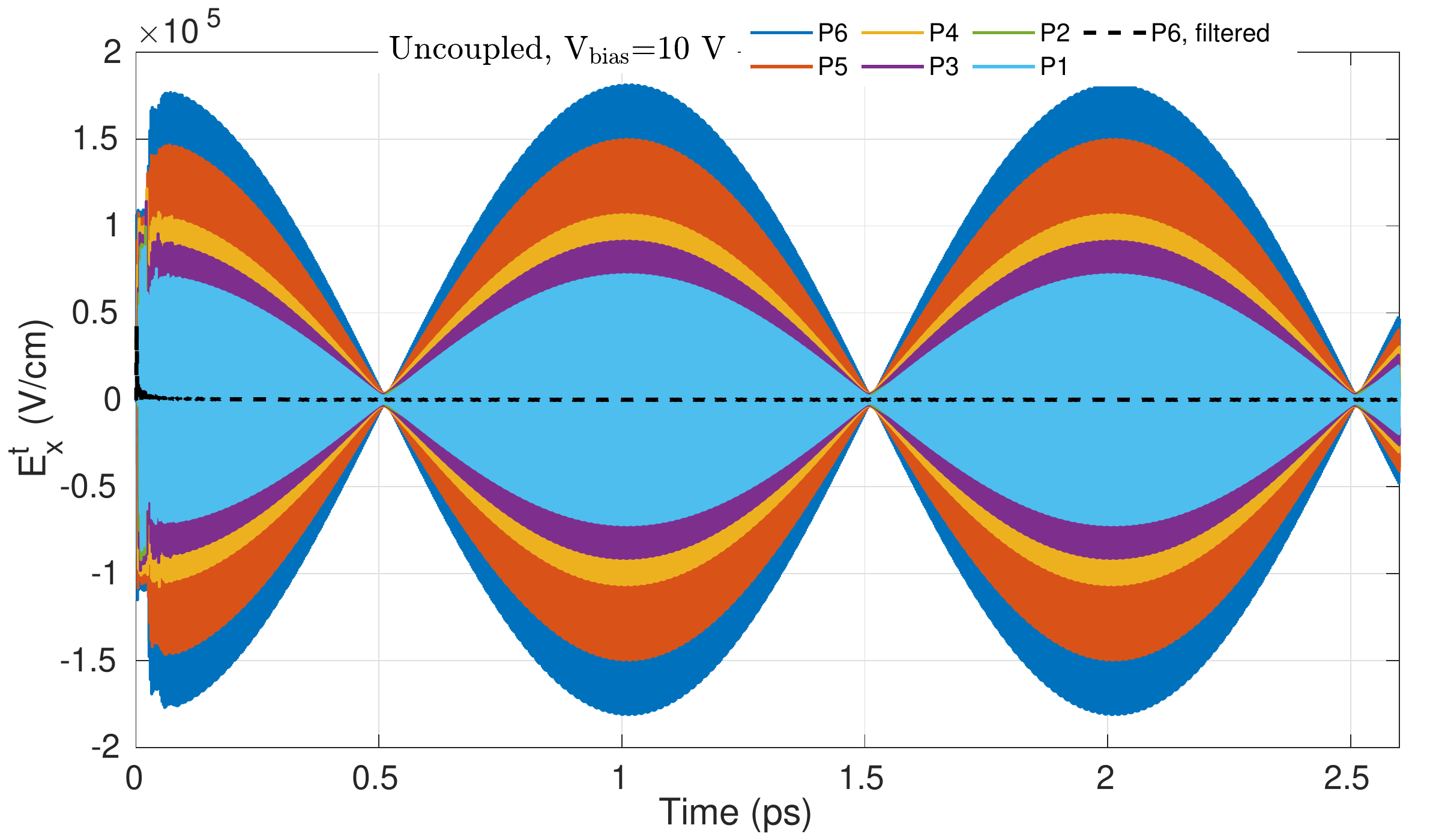}}\\
	\vspace{-0.28cm}
	\subfloat[\label{Fig8Extc}]{\includegraphics[width=0.5\columnwidth]{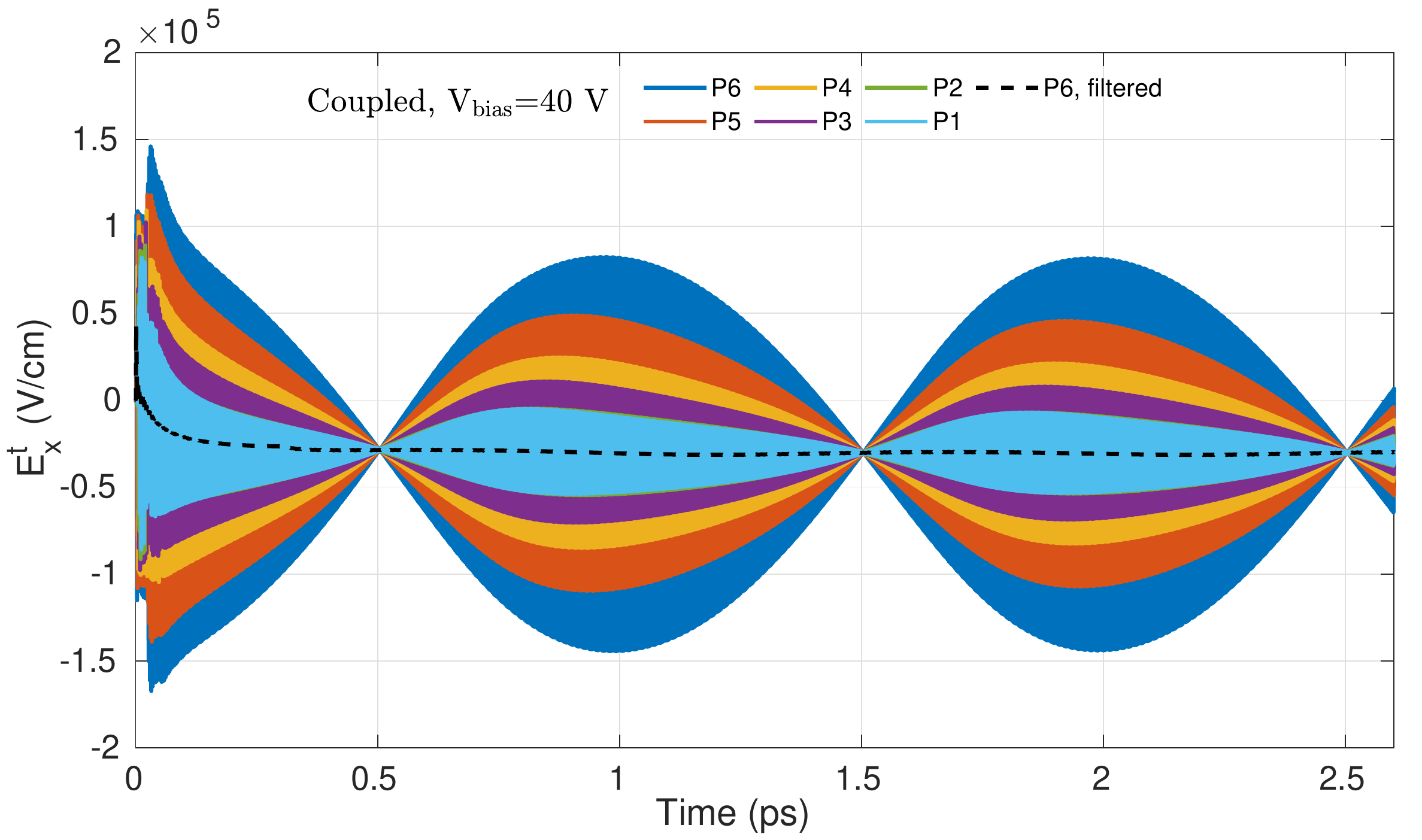}}\\
	\vspace{-0.28cm}
	\subfloat[\label{Fig8Extd}]{\includegraphics[width=0.5\columnwidth]{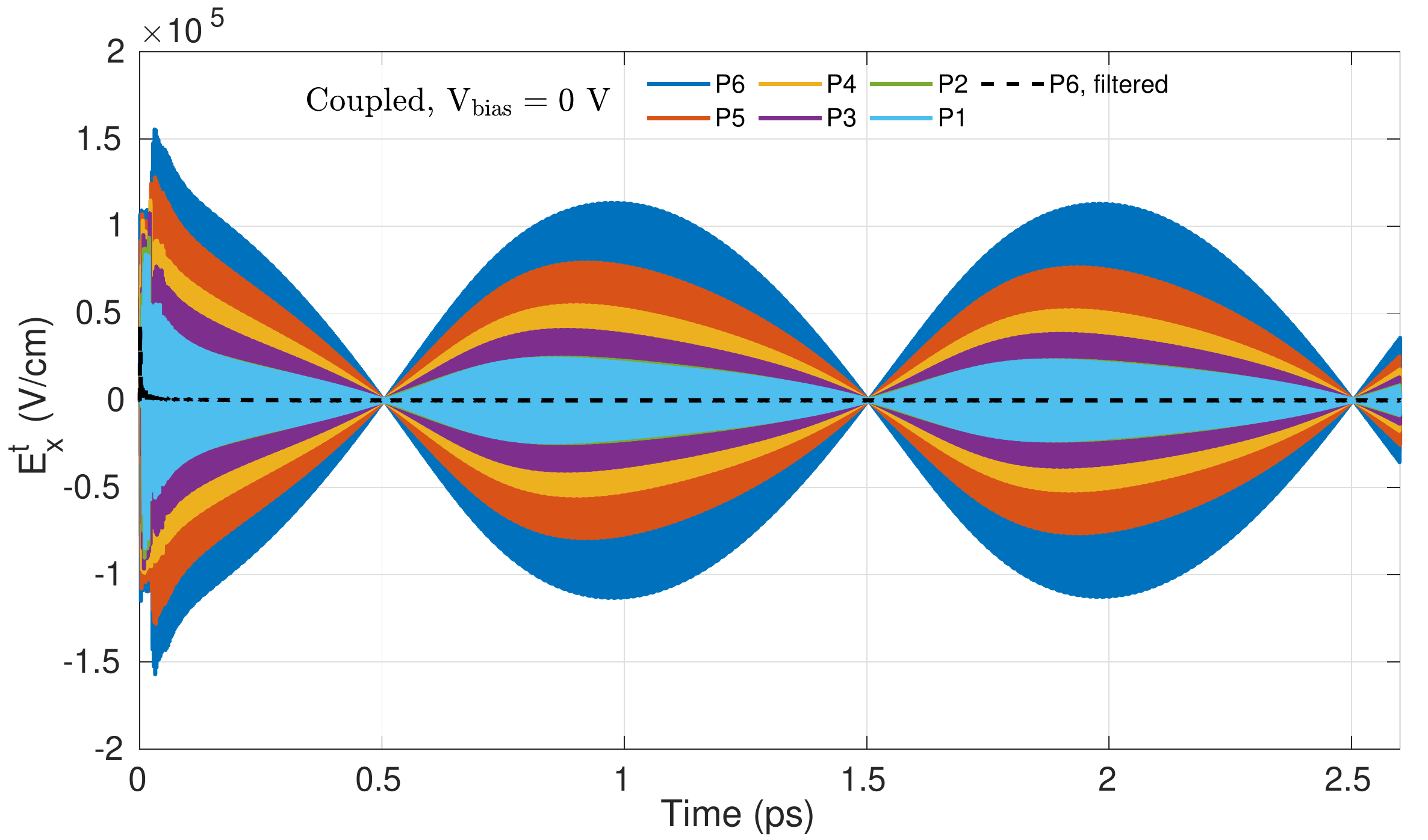}}
	\caption{$E_x^t$ recorded at different probes for {\color{black} $S_\mathrm{pump}=8.3\times 10^{10}$} $\mathrm{mW/cm^2}$ (a) by the coupled simulation with $V_{\mathrm{bias}}=10 \, \mathrm{V}$, (b) by the uncoupled simulation with $V_{\mathrm{bias}}=10 \, \mathrm{V}$, (c) by the coupled simulation with $V_{\mathrm{bias}}=40 \, \mathrm{V}$, and (d) by the coupled simulation with $V_{\mathrm{bias}}=0$. The dashed line represents $E_x^t$ recorded at P6 but smoothened by a low pass filter.}
	\label{Fig8Ext}
\end{figure}

Fig.~\ref{Fig8Ext} plots $E_x^t$ recorded at all six probes versus time for $S_\mathrm{pump}=8.3\times 10^{10} \, \mathrm{mW/cm^2}$ and different values of $V_{\mathrm{bias}}$. The results obtained by the coupled simulation and the uncoupled simulation show several differences:
\begin{itemize}[wide = 0pt]
	\item[(a)] $E_x^t$ computed by the coupled simulation at all probes shows a sudden drop at around $0.1 \, \mathrm{ps}$ [Fig.~\ref{Fig8Ext}(a), (c), and (d)]. After that, the peak values stay smaller than those obtained by the uncoupled simulation [compare Fig.~\ref{Fig8Ext}(a) to Fig.~\ref{Fig8Ext}(b)]. This indicates that in the coupled simulation a smaller level of optical EM field energy can enter the device after a high level of carrier density (or current density) is built up. In other words, the optical EM field energy is ``screened'' when the time-dependent photoconductivity reaches a high level. 
	\item[(b)] The decay observed in $E_x^t$ while moving from the probes closer to the top interface to those located deeper in the device is faster in the coupled simulation [compare lines with different colors at a specific time point between Fig.~\ref{Fig8Ext}(a) and (b)]. This observation agrees with the snapshots shown in Fig.~\ref{Fig7HNe} and is also demonstrated more clearly in the frequency-domain analysis described below. In the coupled simulation, the photoconductivity not only increases the reflection [i.e., does not ``allow’’ the EM fields to enter the active region of the device – see (a)] but also effectively increases the absorption.
	\item[(c)] In Fig.~\ref{Fig8Ext}, the dashed lines represents $E_x^t$ recorded at probe P6 after a low pass filter is applied (by averaging over a sliding window of length $0.25 \, \mathrm{ps}$). In the coupled simulation, the averaged $E_x^t$ oscillates between $-7920 \, \mathrm{V/cm}$ and $-7350 \, \mathrm{V/cm}$ after $1 \, \mathrm{ps}$ [Fig.~\ref{Fig8Ext} (a)] while it is zero in the uncoupled case [Fig.~\ref{Fig8Ext} (b)]. This low frequency field results from the DD currents $\mathbf{J}^t_e(\mathbf{r},t)$ and $\mathbf{J}^t_h(\mathbf{r},t)$ on the right hand side of the Maxwell equation in~\eqref{t_E} and is the reason for the radiation-field screening as analyzed in~\cite{Darrow1992}. Note that the amplitude of the averaged $E_x^t$ is proportional to $V_{\mathrm{bias}}$ [Fig.~\ref{Fig8Ext} (c)] and becomes zero when $V_{\mathrm{bias}}=0$ [Fig.~\ref{Fig8Ext} (d)].
\end{itemize}

\begin{figure}[!t]
	\centering
	\subfloat[\label{Fig9ExfJxfa}]{\includegraphics[width=0.5\columnwidth]{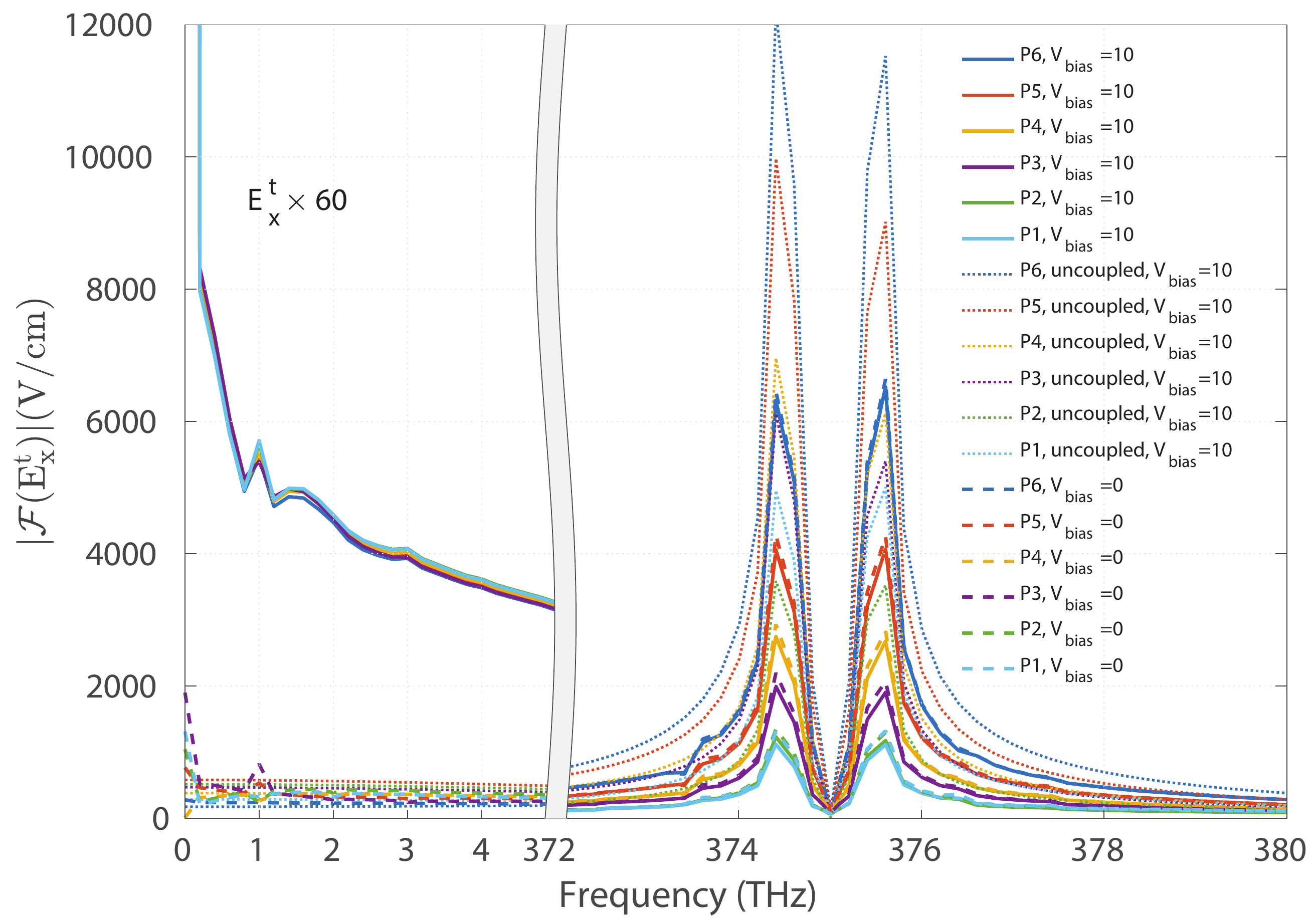}}\\
	\vspace{-0.4cm}
	\subfloat[\label{Fig9ExfJxfb}]{\includegraphics[width=0.5\columnwidth]{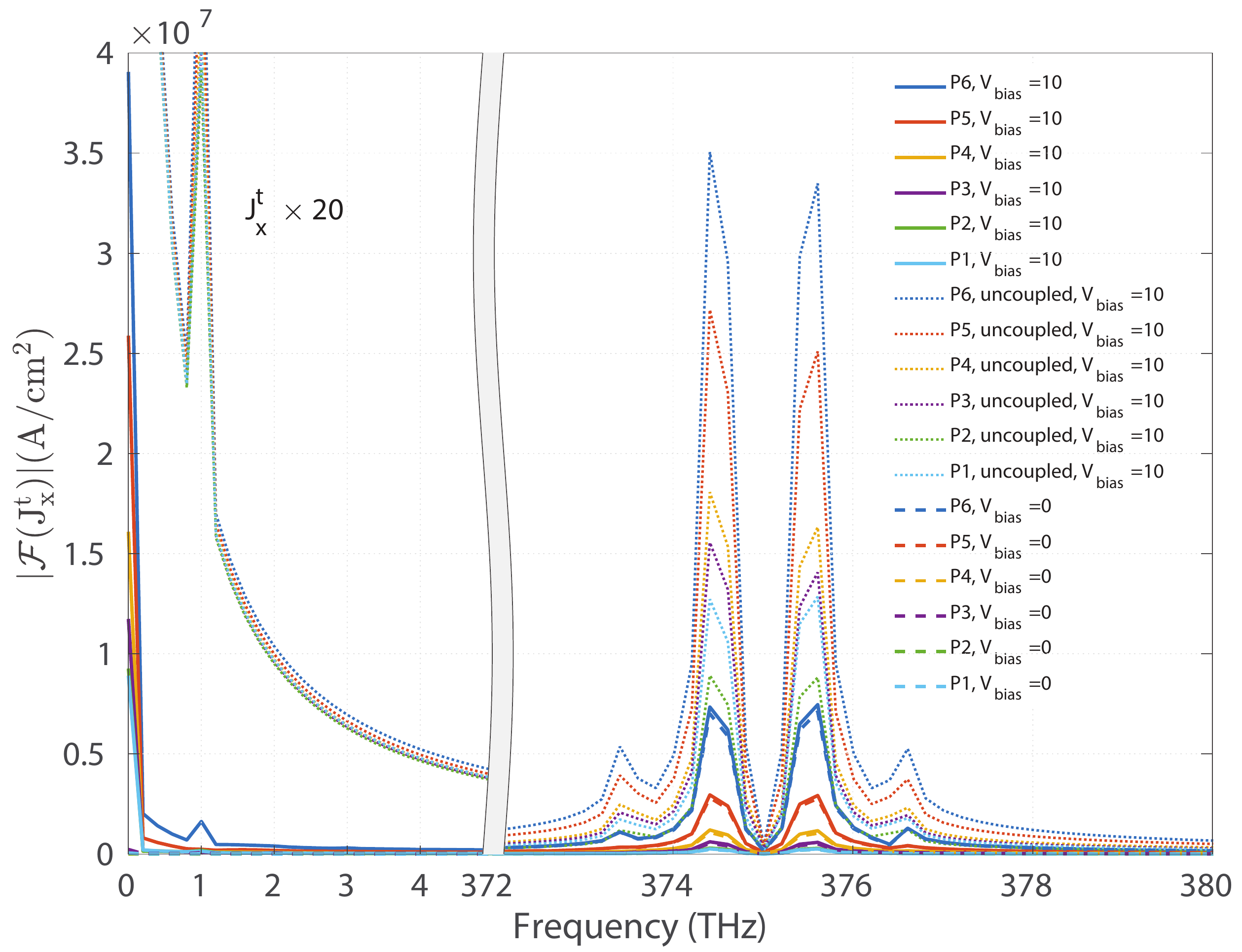}}\\
	\vspace{-0.4cm}
	\subfloat[\label{Fig9Exfc}]{\includegraphics[width=0.249\columnwidth]{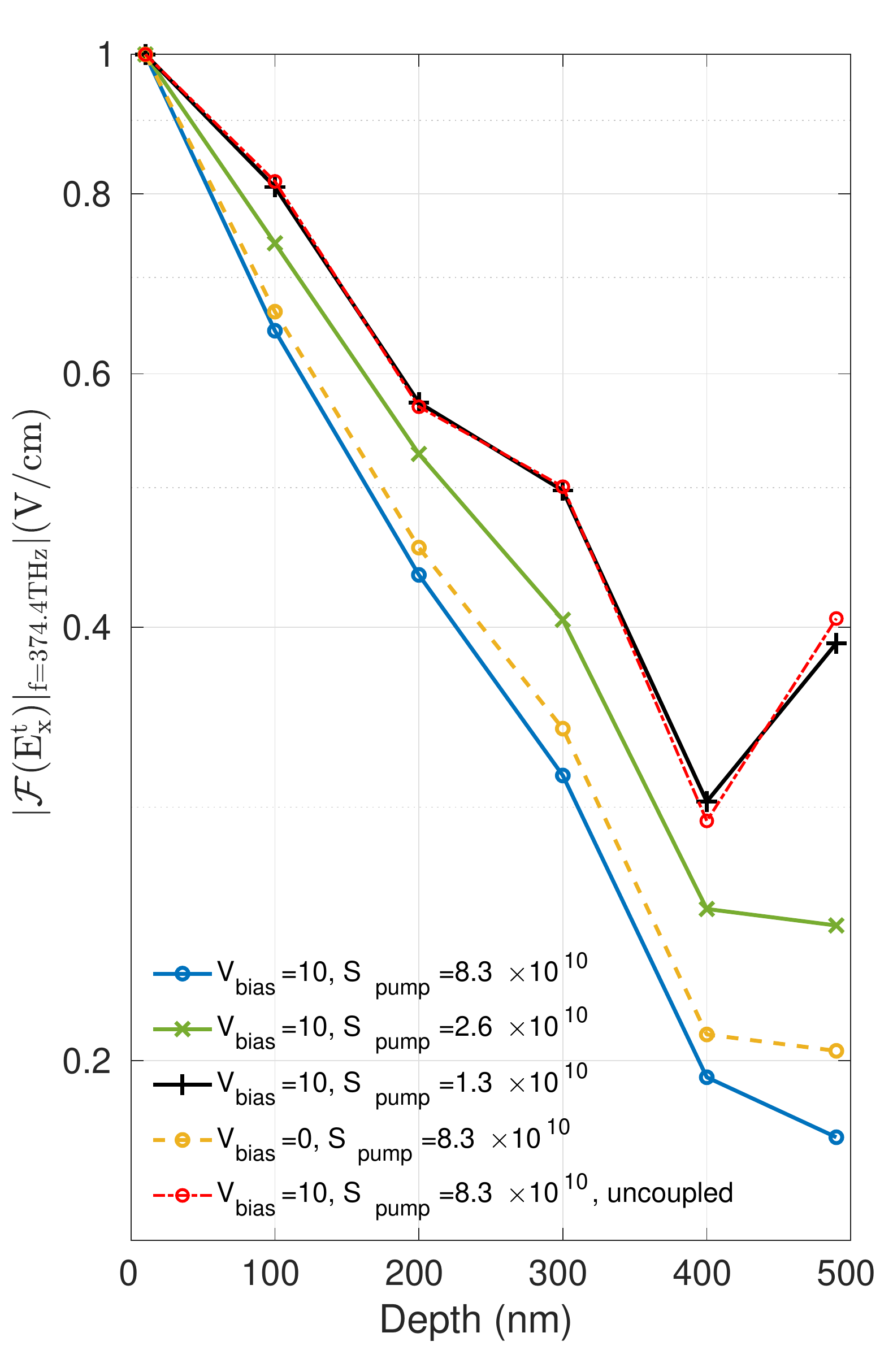}}
	\subfloat[\label{Fig9Jxfd}]{\includegraphics[width=0.249\columnwidth]{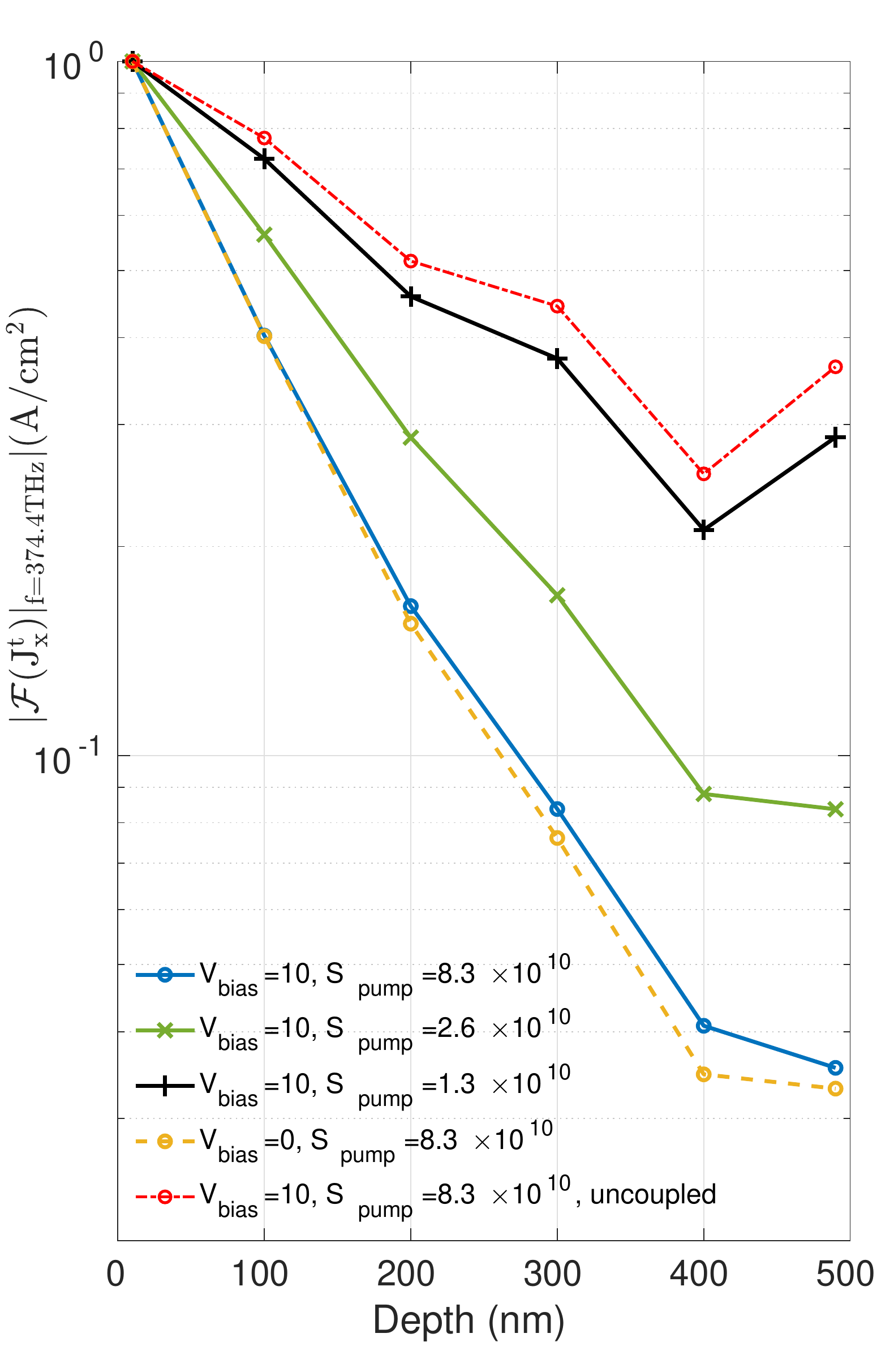}}
	\caption{ (a) $|\mathcal{F}(E_x^t)|$ and (b) $|\mathcal{F}(J_x^t)|$ recorded at different probes. (c) Values of $|\mathcal{F}(E_x^t)|$ and $|\mathcal{F}(J_x^t)|$ at $374.4 \, \mathrm{THz}$ versus depth (along the $-z$ direction) for different values of $S_\mathrm{pump}$. {\color{black}$V_{\mathrm{bias}}$ and $S_{\mathrm{pump}}$ are in units of $\mathrm{V}$ and $\mathrm{mW/cm^2}$, respectively.}}
	\label{Fig9ExfJxf}
	\vspace{-0.3cm}
\end{figure}

To confirm the above observations, the recorded $E_x^t$ and $J_x^t$ are analyzed in the Fourier domain. Fig.~\ref{Fig9ExfJxf} (a) and (b) plot $|\mathcal{F}(E_x^t)|$ and $|\mathcal{F}(J_x^t)|$ versus frequency, respectively. From Fig.~\ref{Fig9ExfJxf} (a), one can see that at the lower frequencies $\mathcal{F}(E_x^t)$ appears in the coupled simulation only when $V_{\mathrm{bias}} \neq 0$. Similarly, as shown in Fig.~\ref{Fig9ExfJxf} (b), at the same frequencies, $\mathcal{F}(J_x^t)$, which is responsible for the THz output of the PCDs, also appears only when $V_{\mathrm{bias}} \neq 0$. Because of the saturation, $\mathcal{F}(E_x^t)$ and $\mathcal{F}(J_x^t)$ are much smaller in the coupled simulation. Near optical frequencies, $\mathcal{F}(E_x^t)$ and $\mathcal{F}(J_x^t)$ are weaker in the coupled simulation. Fig.~\ref{Fig9ExfJxf} (c) plots $|\mathcal{F}(E_x^t)|$ at $374.4 \, \mathrm{THz}$ versus the depth (along the $-z$ direction). Note that, to clearly show the decay rate, each curve is normalized by its value at the top interface ($z = 490$ nm). Clearly, in the coupled simulation, $|\mathcal{F}(E_x^t)|$ at optical frequencies decays faster as $S_\mathrm{pump}$ is increased. This is exactly the optical-field screening effect, i.e., the photoconductivity increases the absorption. Moreover, in the coupled simulation, $|\mathcal{F}(E_x^t)|$ behaves similarly under $V_{\mathrm{bias}}=0$ and $V_{\mathrm{bias}} = 10 \, \mathrm{V}$. Fig.~\ref{Fig9ExfJxf} (d) plots $|\mathcal{F}(J_x^t)|$ at $374.4\,\mathrm{THz}$ versus the depth (along the $-z$ direction), showing that the current density behaves similarly to the electric field.

The above results suggest that the optical-field screening is the dominant mechanism that causes the saturation at the output of the PCD. However, note that it is not trivial to separate the effects of the optical- and radiation-field screening under a bias voltage since both the low-frequency radiation field and the photoconductivity are proportional to the carrier density~\cite{Hu1990, Darrow1992, Benicewicz1994, Siebert2004} and both effects give negative feedback, e.g., a larger density leads to a stronger negative radiation field that effectively reduces the bias electric field, and a larger photoconductivity also reduces the optical power that is allowed to enter the device. To further compare their effects on the saturation, more studies that analyze the transient response the PCDs under different excitation signals might be carried out.

\begin{figure}[!t]
	\centering
	\subfloat[\label{Fig10sigmaa}]{\includegraphics[width=0.6\columnwidth]{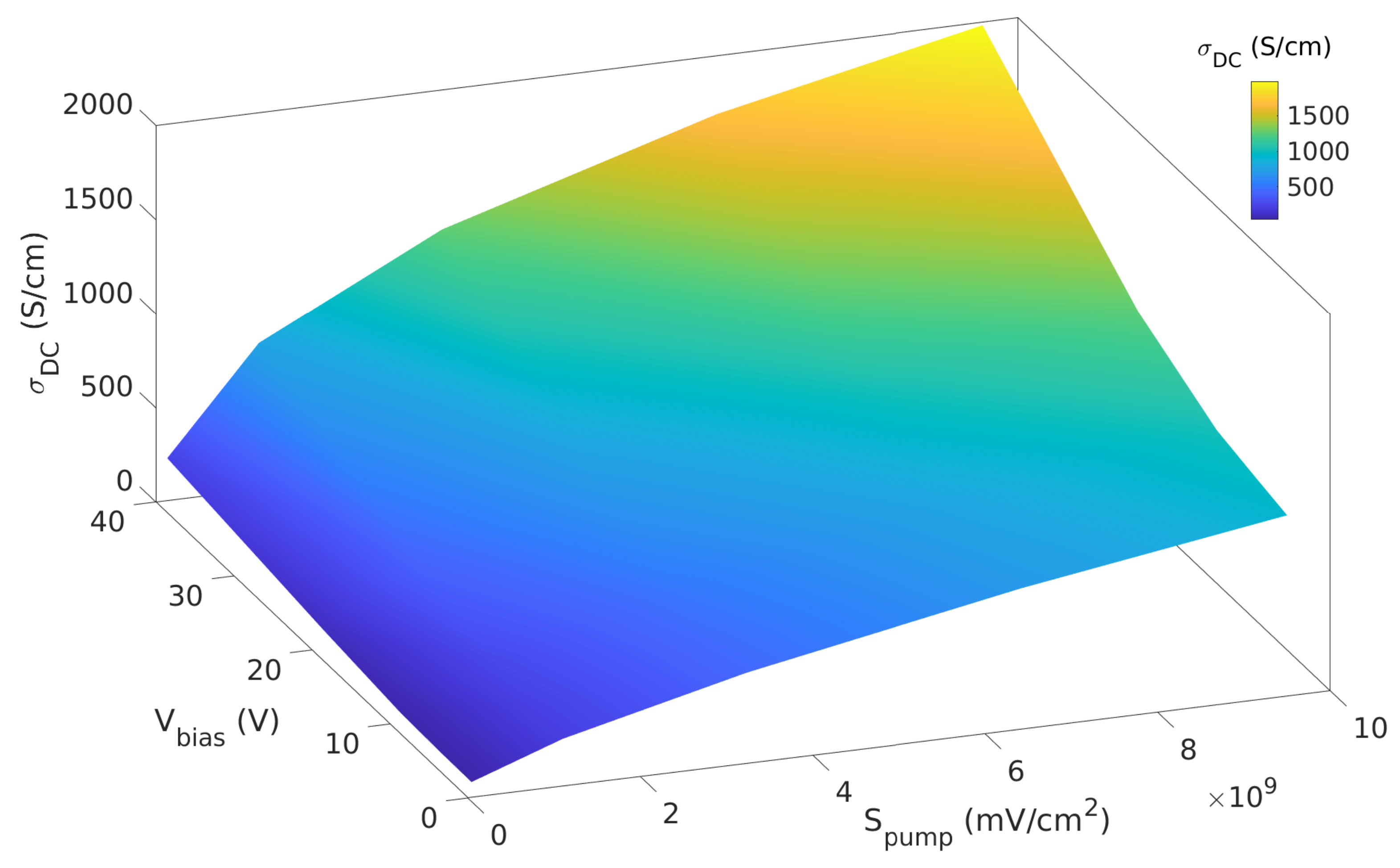}}\\
	\vspace{-0.2cm}
	\subfloat[\label{Fig10sigmav}]{\includegraphics[width=0.6\columnwidth]{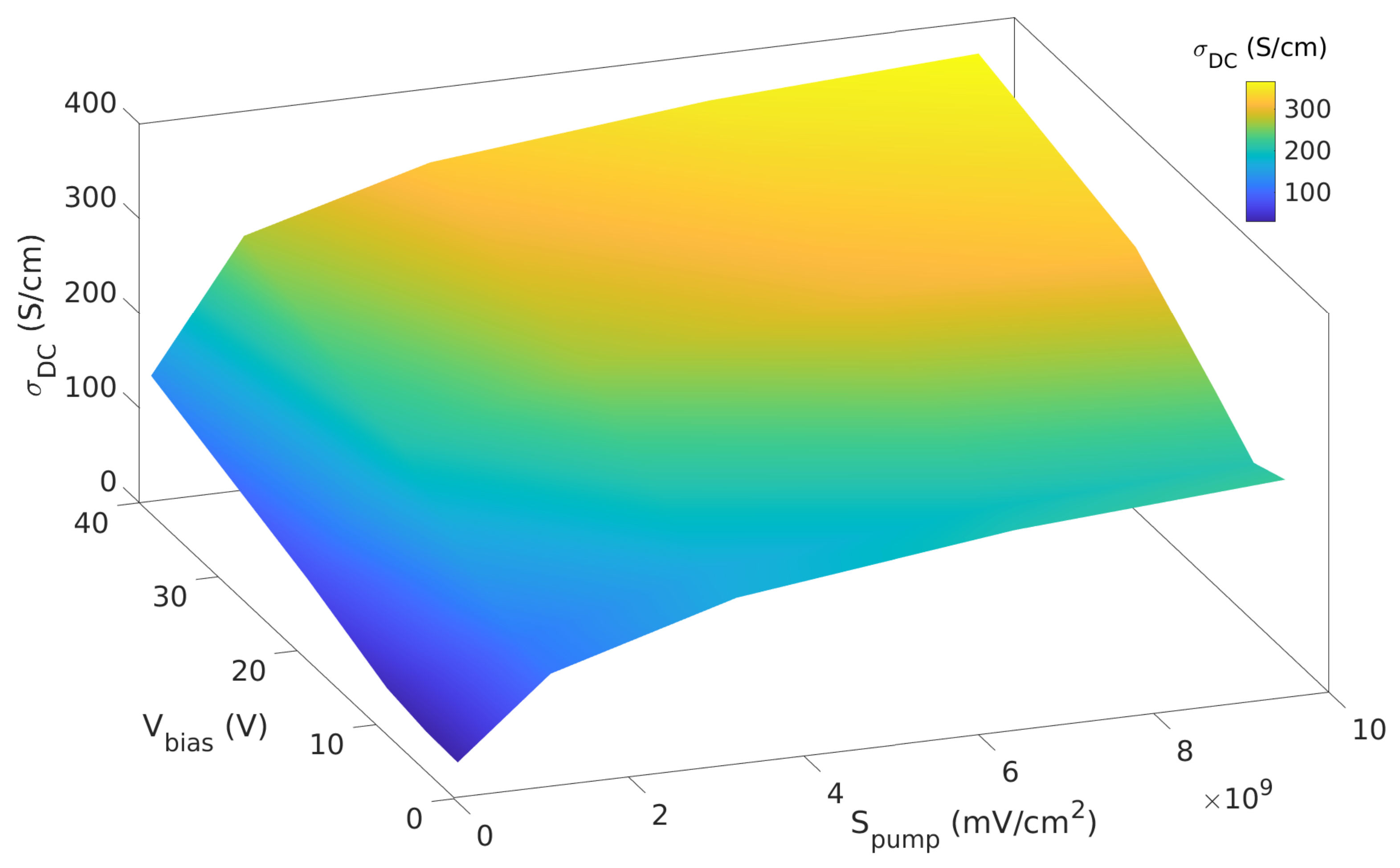}}
	\caption{$\sigma_{DC}$ as a function of $V_{\mathrm{bias}}$ and $S_\mathrm{pump}$ at probes (a) P6 and (b) P1.}
	\label{Fig10sigma}
	\vspace{-0.2cm}
\end{figure}

Clearly, the photoconductivity is time-dependent. This also means that it is frequency-dependent~\cite{Ulbricht2011RMP}. The DC component of the conductivity can be defined as 
\begin{equation*}
\sigma_{DC} = [\mathcal{F}(J_x^t)|_{f=0}+J_x^s]/[\mathcal{F}(E_x^t)|_{f=0}+E_x^s]
\end{equation*}
where $E_x^s$ and $J_x^s$ are the $x$ components of the steady-state electric field and current density. Note that $\mathcal{F}(J_x^t)|_{f=0}$ and $\mathcal{F}(E_x^t)|_{f=0}$ are real and therefore $\sigma_{DC}$ is real. The saturation behavior indicates that the photoconductivity depends not only on $S_\mathrm{pump}$, but also on $V_{\mathrm{bias}}$. Moreover, it also depends on the time signature of the pump, e.g., the repetition rate of a pulse affects the dynamic response. Here, the time signature is same as the one in the above examples. Fig.~\ref{Fig10sigma} plots $\sigma_{DC}$ computed at probes P6 and P1 as a function of $S_\mathrm{pump}$ and $V_{\mathrm{bias}}$. Since the carrier density is much higher closer to the top interface, $\sigma_{DC}$ is larger at probe P6. At probe P6, one can see that the saturation at high levels of optical pump for a given value of $V_\mathrm{bias}$. No saturation along $V_\mathrm{bias}$ is observed for given value of $S_\mathrm{pump}$. This agrees with the observation reported in~\cite{Darrow1992, Benicewicz1994, Liu2003, Gobel2011}. At probe P1, $\sigma_{DC}$ saturates with increasing $S_\mathrm{pump}$ or $V_\mathrm{bias}$ since the conductivity is much lower closer to the bottom interface.

\section{Conclusion}

The saturation behavior of the PCD output under high optical pump power levels and the underlying screening effects are investigated using an efficient fully- and rigorously-coupled multiphysics approach. Since this approach models the two-way nonlinear coupling between the EM fields and the carriers, it enables the numerical characterization of the screening effects. In addition, the efficient unit-cell model permits generation of numerical results from a large number of simulations in a reasonable time.

Excellent agreement is found between the numerical results and the experimental data. Numerical results generated using a large number of simulations clearly show the effects optical- and radiation-field screening. The impacts of these screening effects are compared by controlling the coupling mechanism in the multiphysics model. It is demonstrated that the saturation of the output is observed only when the photocurrent is accounted for in the Maxwell equations (i.e., when the carrier effects are coupled back onto the EM fields). The low-frequency EM fields generated by this photocurrent dampens the bias electric field (radiation-field screening). Additionally, a higher carrier density level increases the effective photoconductivity, which limits the penetration of the optical EM fields into the active region of the device (optical-field screening). Without the coupling of the photocurrent to the Maxwell equations, the EM field interactions do not ``see'' this increasing photoconductivity. Finally, it is found that the optical-field screening plays a more dominant role in the saturation of the PCD output under high optical pump power levels.

{\color{black} In recent years, nanostructures are extensively used for improving the PCD response by enhancing the optical EM fields that interact with the active region of the device. From the results provided in this work, one can expect that the high carrier densities generated by the enhanced EM optical fields near the nanostructures would block optical EM fields from penetrating into the active region, possibly leading to more pronounced optical-field screening effect. Therefore, the nanostructures should be designed not only to enhance the EM optical fields locally but also to guide them more effectively into the active region. From this perspective, three dimensional nanostructures that are etched into (rather than onto) the active region are promising candidates.}

\section*{Acknowledgment}
This work was supported by the King Abdullah University of Science and Technology Office of Sponsored Research under Award 2016-CRG5-2953 {\color{black}and Award 2019-CRG8-4056}. The authors would like to thank the KAUST Supercomputing Laboratory (KSL) for providing the required computational resources.

\bibliographystyle{IEEEtran}

\begin{thebibliography}{10}
	\providecommand{\url}[1]{#1}
	\csname url@samestyle\endcsname
	\providecommand{\newblock}{\relax}
	\providecommand{\bibinfo}[2]{#2}
	\providecommand{\BIBentrySTDinterwordspacing}{\spaceskip=0pt\relax}
	\providecommand{\BIBentryALTinterwordstretchfactor}{4}
	\providecommand{\BIBentryALTinterwordspacing}{\spaceskip=\fontdimen2\font plus
		\BIBentryALTinterwordstretchfactor\fontdimen3\font minus
		\fontdimen4\font\relax}
	\providecommand{\BIBforeignlanguage}[2]{{%
			\expandafter\ifx\csname l@#1\endcsname\relax
			\typeout{** WARNING: IEEEtran.bst: No hyphenation pattern has been}%
			\typeout{** loaded for the language `#1'. Using the pattern for}%
			\typeout{** the default language instead.}%
			\else
			\language=\csname l@#1\endcsname
			\fi
			#2}}
	\providecommand{\BIBdecl}{\relax}
	\BIBdecl
	
	\bibitem{Lee2018}
	E.~S. {Lee}, K.~{Moon}, I.~{Lee}, H.~{Kim}, D.~W. {Park}, J.~{Park}, D.~H.
	{Lee}, S.~{Han}, N.~{Kim}, and K.~H. {Park}, ``Semiconductor-based terahertz
	photonics for industrial applications,'' \emph{J. Lightwave Technol.},
	vol.~36, no.~2, pp. 274--283, 2018.
	
	\bibitem{Saeedkia2008}
	D.~{Saeedkia} and S.~{Safavi-Naeini}, ``Terahertz photonics: {O}ptoelectronic
	techniques for generation and detection of terahertz waves,'' \emph{J.
		Lightwave Technol.}, vol.~26, no.~15, pp. 2409--2423, 2008.
	
	\bibitem{Lepeshov2017review}
	S.~Lepeshov, A.~Gorodetsky, A.~Krasnok, E.~Rafailov, and P.~Belov,
	``Enhancement of terahertz photoconductive antenna operation by optical
	nanoantennas,'' \emph{Laser Photonics Rev.}, vol.~11, no.~1, p. 1600199,
	2017.
	
	\bibitem{Burford2017review}
	N.~M. Burford and M.~O. El-Shenawee, ``Review of terahertz photoconductive
	antenna technology,'' \emph{Opt. Eng.}, vol.~56, no.~1, p. 010901, 2017.
	
	\bibitem{Kang2018review}
	J.-H. Kang, D.-S. Kim, and M.~Seo, ``Terahertz wave interaction with metallic
	nanostructures,'' \emph{Nanophotonics}, vol.~7, no.~5, pp. 763--793, 2018.
	
	\bibitem{Yardimci2018review}
	N.~T. Yardimci and M.~Jarrahi, ``Nanostructure-enhanced photoconductive
	terahertz emission and detection,'' \emph{Small}, vol.~14, no.~44, p.
	1802437, 2018.
	
	\bibitem{Yachmenev2019review}
	A.~E. Yachmenev, D.~V. Lavrukhin, I.~A. Glinskiy, N.~V. Zenchenko, Y.~G.
	Goncharov, I.~E. Spektor, R.~A. Khabibullin, T.~Otsuji, and D.~S. Ponomarev,
	``Metallic and dielectric metasurfaces in photoconductive terahertz devices:
	{A} review,'' \emph{Opt. Eng.}, vol.~59, no.~6, p. 061608, 2019.
	
	\bibitem{Ramer2015}
	J.~{Ramer} and G.~{von Freymann}, ``A terahertz time-domain spectroscopy-based
	network analyzer,'' \emph{J. Lightwave Technol.}, vol.~33, no.~2, pp.
	403--407, 2015.
	
	\bibitem{Torkaman2018}
	P.~{Torkaman}, S.~{Darbari}, and M.~J. {Mohammad-Zamani}, ``Design and
	simulation of a piezotronic gan-based pulsed {THz} emitter,'' \emph{J.
		Lightwave Technol.}, vol.~36, no.~17, pp. 3645--3651, 2018.
	
	\bibitem{Moon2021}
	S.~R. {Moon}, M.~{Sung}, J.~K. {Lee}, and S.~H. {Cho}, ``Cost-effective
	photonics-based {THz} wireless transmission using pam-n signals in the 0.3
	{THz} band,'' \emph{J. Lightwave Technol.}, vol.~39, no.~2, pp. 357--362,
	2021.
	
	\bibitem{Berry2013}
	C.~W. Berry, N.~Wang, M.~R. Hashemi, M.~Unlu, and M.~Jarrahi, ``Significant
	performance enhancement in photoconductive terahertz optoelectronics by
	incorporating plasmonic contact electrodes,'' \emph{Nat. Commun.}, vol.~4, p.
	1622, 2013.
	
	\bibitem{Yang2014}
	S.-H. Yang, M.~R. Hashemi, C.~W. Berry, and M.~Jarrahi, ``7.5\%
	optical-to-terahertz conversion efficiency offered by photoconductive
	emitters with three-dimensional plasmonic contact electrodes,'' \emph{IEEE
		Trans. THz Sci. Technol.}, vol.~4, no.~5, pp. 575--581, 2014.
	
	\bibitem{Darrow1992}
	J.~T. {Darrow}, X.~C. {Zhang}, D.~H. {Auston}, and J.~D. {Morse}, ``Saturation
	properties of large-aperture photoconducting antennas,'' \emph{IEEE J.
		Quantum Electron.}, vol.~28, no.~6, pp. 1607--1616, 1992.
	
	\bibitem{Benicewicz1994}
	P.~K. Benicewicz and A.~J. Taylor, ``Scaling of terahertz radiation from
	large-aperture biased inp photoconductors,'' \emph{Opt. Lett.}, vol.~18,
	no.~16, pp. 1332--1334, 1993.
	
	\bibitem{Siebert2004}
	K.~J. Siebert, A.~Lisauskas, T.~Löffler, and H.~G. Roskos, ``Field screening
	in low-temperature-grown {{GaAs}} photoconductive antennas,'' \emph{Jpn. J.
		App. Phy.}, vol.~43, no.~3, pp. 1038--1043, 2004.
	
	\bibitem{Kim2006}
	D.~S. Kim and D.~S. Citrin, ``Coulomb and radiation screening in
	photoconductive terahertz sources,'' \emph{Appl. Phys. Lett.}, vol.~88,
	no.~16, p. 161117, 2006.
	
	\bibitem{Rodriguez1996}
	G.~Rodriguez and A.~J. Taylor, ``Screening of the bias field in terahertz
	generation from photoconductors,'' \emph{Opt. Lett.}, vol.~21, no.~14, pp.
	1046--1048, 1996.
	
	\bibitem{Tani1997}
	M.~Tani, S.~Matsuura, K.~Sakai, and S.-i. Nakashima, ``Emission characteristics
	of photoconductive antennas based on low-temperature-grown {GaAs} and
	semi-insulating {GaAs},'' \emph{Appl. Opt.}, vol.~36, no.~30, pp. 7853--7859,
	1997.
	
	\bibitem{Loata2007}
	G.~C. Loata, M.~D. Thomson, T.~Loffler, and H.~G. Roskos, ``Radiation field
	screening in photoconductive antennae studied via pulsed terahertz emission
	spectroscopy,'' \emph{Appl. Phys. Lett.}, vol.~91, no.~23, p. 232506, 2007.
	
	\bibitem{Chou2013}
	R.-H. Chou, C.-S. Yang, and C.-L. Pan, ``Effects of pump pulse propagation and
	spatial distribution of bias fields on terahertz generation from
	photoconductive antennas,'' \emph{J. Appl. Phys.}, vol. 114, no.~4, p.
	043108, 2013.
	
	\bibitem{Burford2016}
	N.~Burford and M.~El-Shenawee, ``Computational modeling of plasmonic thin-film
	terahertz photoconductive antennas,'' \emph{J. Opt. Soc. Am. B}, vol.~33,
	no.~4, pp. 748--759, 2016.
	
	\bibitem{Gagnon2016}
	E.~Gagnon, N.~K. Owusu, and A.~L. Lytle, ``Time evolution of the coulomb
	screening effects on terahertz generation at the surface of inas,'' \emph{J.
		Opt. Soc. Am. B}, vol.~33, no.~3, pp. 367--372, 2016.
	
	\bibitem{Khorshidi2016}
	M.~Khorshidi and G.~Dadashzadeh, ``Plasmonic photoconductive antennas with
	rectangular and stepped rods: {A} theoretical analysis,'' \emph{J. Opt. Soc.
		Am. B}, vol.~33, no.~12, pp. 2502--2511, 2016.
	
	\bibitem{Burford2018}
	N.~M. {Burford}, M.~J. {Evans}, and M.~O. {El-Shenawee}, ``Plasmonic nanodisk
	thin-film terahertz photoconductive antenna,'' \emph{IEEE Trans. THz Sci.
		Technol.}, vol.~8, no.~2, pp. 237--247, 2018.
	
	\bibitem{Hu1990}
	B.~B. Hu, J.~T. Darrow, X.~Zhang, D.~H. Auston, and P.~R. Smith, ``Optically
	steerable photoconducting antennas,'' \emph{Appl. Phys. Lett.}, vol.~56,
	no.~10, pp. 886--888, 1990.
	
	\bibitem{Sirbu2005}
	M.~{Sirbu}, S.~B.~P. {Lepaul}, and F.~{Aniel}, ``Coupling 3-{D} {M}axwell's and
	{B}oltzmann's equations for analyzing a terahertz photoconductive switch,''
	\emph{IEEE Trans. Microw. Theory Tech.}, vol.~53, no.~9, pp. 2991--2998,
	2005.
	
	\bibitem{Neshat2010}
	M.~Neshat, D.~Saeedkia, L.~Rezaee, and S.~Safavi-Naeini, ``A global approach
	for modeling and analysis of edge-coupled traveling-wave terahertz
	photoconductive sources,'' \emph{IEEE Trans. Microw. Theory Tech.}, vol.~58,
	no.~7, pp. 1952--1966, 2010.
	
	\bibitem{Kirawanich2008}
	P.~Kirawanich, S.~J. Yakura, and N.~E. Islam, ``Study of high-power wideband
	terahertz-pulse generation using integrated high-speed photoconductive
	semiconductor switches,'' \emph{IEEE Trans. Plasma Sci.}, vol.~37, no.~1, pp.
	219--228, 2008.
	
	\bibitem{Khabiri2012}
	M.~Khabiri, M.~Neshat, and S.~Safavi-Naeini, ``Hybrid computational simulation
	and study of continuous wave terahertz photomixers,'' \emph{IEEE Trans. THz
		Sci. Technol.}, vol.~2, no.~6, pp. 605--616, 2012.
	
	\bibitem{Khiabani2013}
	N.~Khiabani, Y.~Huang, Y.-C. Shen, and S.~Boyes, ``Theoretical modeling of a
	photoconductive antenna in a terahertz pulsed system,'' \emph{IEEE Trans.
		Antennas Propag.}, vol.~61, no.~4, pp. 1538--1546, 2013.
	
	\bibitem{Young2014}
	J.~C. Young, D.~Boyd, S.~D. Gedney, T.~Suzuki, and J.~Liu, ``A {DGFETD} port
	formulation for photoconductive antenna analysis,'' \emph{IEEE Antennas
		Wireless Propag. Lett.}, vol.~14, pp. 386--389, 2014.
	
	\bibitem{Moreno2014}
	E.~Moreno, M.~F. Pantoja, S.~G. Garcia, A.~R. Bretones, and R.~G. Martin,
	``Time-domain numerical modeling of {TH}z photoconductive antennas,''
	\emph{IEEE Trans. THz Sci. Technol.}, vol.~4, no.~4, pp. 490--500, 2014.
	
	\bibitem{Mohammad2016nanoslit}
	M.~J. Mohammad-Zamani, M.~Neshat, and M.~K. Moravvej-Farshi, ``Nanoslit cavity
	plasmonic modes and built-in fields enhance the {CW} {TH}z radiation in an
	unbiased antennaless photomixers array,'' \emph{Opt. Lett.}, vol.~41, no.~2,
	pp. 420--423, 2016.
	
	\bibitem{Bashirpour2017}
	M.~Bashirpour, S.~Ghorbani, M.~Kolahdouz, M.~Neshat, M.~Masnadi-Shirazi, and
	H.~Aghababa, ``Significant performance improvement of a terahertz
	photoconductive antenna using a hybrid structure,'' \emph{RSC Advances},
	vol.~7, no.~83, pp. 53\,010--53\,017, 2017.
	
	\bibitem{Burford2017}
	N.~M. Burford, M.~J. Evans, and M.~O. El-Shenawee, ``Plasmonic nanodisk
	thin-film terahertz photoconductive antenna,'' \emph{IEEE Trans. THz Sci.
		Technol.}, vol.~8, no.~2, pp. 237--247, 2017.
	
	\bibitem{Chen2019multiphysics}
	L.~Chen and H.~Bagci, ``Multiphysics simulation of plasmonic photoconductive
	devices using discontinuous {G}alerkin methods,'' \emph{IEEE J. Multiscale
		Multiphys. Comput. Tech.}, vol.~5, pp. 188--200, 2020.
	
	\bibitem{Chen2020steadystate}
	L.~{Chen} and H.~{Bagci}, ``Steady-state simulation of semiconductor devices
	using discontinuous {G}alerkin methods,'' \emph{IEEE Access}, vol.~8, pp.
	16\,203--16\,215, 2020.
	
	\bibitem{Chen2019unitcell}
	------, ``A unit-cell discontinuous {G}alerkin scheme for analyzing plasmonic
	photomixers,'' in \emph{Proc. IEEE Int. Symp. Antennas Propag.}, 2019, pp.
	1069--1070.
	
	\bibitem{Chen2020efficient}
	L.~Chen, K.~Sirenko, and H.~Bagci, ``An efficient discontinuous {G}alerkin
	scheme for simulating terahertz photoconductive devices with periodic
	nanostructures,'' \emph{arXiv preprint arXiv:2006.02141}, 2020.
	
	\bibitem{Vasileska2010}
	D.~Vasileska, S.~M. Goodnick, and G.~Klimeck, \emph{Computational Electronics:
		{S}emiclassical and quantum device modeling and simulation}.\hskip 1em plus
	0.5em minus 0.4em\relax Boca Raton, FL, USA: CRC press, 2010.
	
	\bibitem{Chen2019discontinuous}
	L.~Chen and H.~Bagci, ``A discontinuous {G}alerkin framework for multiphysics
	simulation of photoconductive devices,'' in \emph{Proc. Int. Appl. Comput.
		Electromagn. Symp.}, 2019, pp. 1--2.
	
	\bibitem{Chen2021generation}
	------, ``Material absorption-based carrier generation model for modeling
	optoelectronic devices,'' \emph{arXiv preprint arXiv:2102.06702}, 2021.
	
	\bibitem{Cockburn1998}
	B.~Cockburn and C.-W. Shu, ``The local discontinuous {G}alerkin method for
	time-dependent convection-diffusion systems,'' \emph{SIAM J. Numer. Anal.},
	vol.~35, no.~6, pp. 2440--2463, 1998.
	
	\bibitem{Castillo2000}
	P.~Castillo, B.~Cockburn, I.~Perugia, and D.~Schotzau, ``An a priori error
	analysis of the local discontinuous {G}alerkin method for elliptic
	problems,'' \emph{SIAM J. Numer. Anal.}, vol.~38, no.~5, pp. 1676--1706,
	2000.
	
	\bibitem{Shu2016}
	C.-W. {Shu}, \emph{Discontinuous {G}alerkin methods for time-dependent
		convection dominated problems: {B}asics, recent developments and comparison
		with other methods}.\hskip 1em plus 0.5em minus 0.4em\relax Cham: Springer
	International Publishing, 2016, pp. 371--399.
	
	\bibitem{Chen2020float}
	L.~{Chen}, M.~{Dong}, and H.~{Bagci}, ``Modeling floating potential conductors
	using discontinuous {G}alerkin method,'' \emph{IEEE Access}, vol.~8, pp.
	7531--7538, 2020.
	
	\bibitem{Chen2020hybridizable}
	L.~Chen, M.~Dong, P.~Li, and H.~Bagci, ``A hybridizable discontinuous
	{G}alerkin method for simulation of electrostatic problems with floating
	potential conductors,'' \emph{Int. J. Numer. Model.: Electron. Networks,
		Device Fields}, p. e2894, 2020.
	
	\bibitem{Hesthaven2002}
	J.~Hesthaven and T.~Warburton, ``Nodal high-order methods on unstructured
	grids: {I}. time-domain solution of {M}axwell's equations,'' \emph{J. Comput.
		Phys.}, vol. 181, no.~1, pp. 186 -- 221, 2002.
	
	\bibitem{Lu2004}
	T.~Lu, P.~Zhang, and W.~Cai, ``Discontinuous {G}alerkin methods for dispersive
	and lossy {M}axwell's equations and {PML} boundary conditions,'' \emph{J.
		Comput. Phys.}, vol. 200, no.~2, pp. 549--580, 2004.
	
	\bibitem{Fezoui2005}
	L.~Fezoui, S.~Lanteri, S.~Lohrengel, and S.~Piperno, ``Convergence and
	stability of a discontinuous {G}alerkin time-domain method for the 3{D}
	heterogeneous {M}axwell equations on unstructured meshes,'' \emph{ESAIM-Math.
		Model. Numer. Anal.}, vol.~39, no.~6, pp. 1149--1176, 2005.
	
	\bibitem{Hesthaven2008}
	J.~Hesthaven and T.~Warburton, \emph{Nodal Discontinuous {G}alerkin Methods:
		Algorithms, Analysis, and Applications}.\hskip 1em plus 0.5em minus
	0.4em\relax NY, USA: Springer, 2008.
	
	\bibitem{Niegemann2009}
	J.~Niegemann, M.~Konig, K.~Stannigel, and K.~Busch, ``Higher-order time-domain
	methods for the analysis of nano-photonic systems,'' \emph{Photonic.
		Nanostruct.}, vol.~7, no.~1, pp. 2--11, 2009.
	
	\bibitem{Gedney2009}
	S.~D. Gedney, C.~Luo, J.~A. Roden, R.~D. Crawford, B.~Guernsey, J.~A. Miller,
	T.~Kramer, and E.~W. Lucas, ``The discontinuous {G}alerkin finite-element
	time-domain method solution of {M}axwell's equations,'' \emph{Appl. Comput.
		Electromagn. Soc. J.}, vol.~24, no.~2, p. 129, 2009.
	
	\bibitem{Liu2012}
	M.~Liu, K.~Sirenko, and H.~Bagci, ``An efficient discontinuous {G}alerkin
	finite element method for highly accurate solution of {M}axwell equations,''
	\emph{IEEE Trans. Antennas Propag.}, vol.~60, no.~8, pp. 3992--3998, 2012.
	
	\bibitem{Sirenko2012}
	K.~Sirenko, M.~Liu, and H.~Bagci, ``Incorporation of exact boundary conditions
	into a discontinuous {G}alerkin finite element method for accurately solving
	{2D} time-dependent {M}axwell equations,'' \emph{IEEE Trans. Antennas
		Propag.}, vol.~61, no.~1, pp. 472--477, 2012.
	
	\bibitem{Chen2012discontinuous}
	J.~Chen and Q.~H. Liu, ``Discontinuous {G}alerkin time-domain methods for
	multiscale electromagnetic simulations: {A} review,'' \emph{Proc. IEEE}, vol.
	101, no.~2, pp. 242--254, 2012.
	
	\bibitem{Li2015IBC}
	P.~{Li}, Y.~{Shi}, L.~J. {Jiang}, and H.~{Bagci}, ``{DGTD} analysis of
	electromagnetic scattering from penetrable conductive objects with {IBC},''
	\emph{IEEE Trans. Antennas Propag.}, vol.~63, no.~12, pp. 5686--5697, 2015.
	
	\bibitem{Li2017dispersive}
	P.~{Li}, L.~J. {Jiang}, and H.~{Bagci}, ``Transient analysis of dispersive
	power-ground plate pairs with arbitrarily shaped antipads by the {DGTD}
	method with wave port excitation,'' \emph{IEEE Trans. Electromagn. Compat.},
	vol.~59, no.~1, pp. 172--183, 2017.
	
	\bibitem{Sirenko2018}
	K.~{Sirenko}, Y.~{Sirenko}, and H.~{Bagci}, ``Exact absorbing boundary
	conditions for periodic three-dimensional structures: {D}erivation and
	implementation in discontinuous {G}alerkin time-domain method,'' \emph{IEEE
		J. Multiscale Multiphys. Comput. Tech.}, vol.~3, pp. 108--120, 2018.
	
	\bibitem{Liu2004}
	Y.~Liu and C.-W. Shu, ``Local discontinuous {G}alerkin methods for moment
	models in device simulations: {F}ormulation and one dimensional results,''
	\emph{J. Comput. Electron.}, vol.~3, no. 3-4, pp. 263--267, 2004.
	
	\bibitem{Liu2016}
	------, ``Analysis of the local discontinuous {G}alerkin method for the
	drift-diffusion model of semiconductor devices,'' \emph{Sci. China Math.},
	vol.~59, no.~1, pp. 115--140, 2016.
	
	\bibitem{Harmon2016}
	M.~Harmon, I.~M. Gamba, and K.~Ren, ``Numerical algorithms based on {G}alerkin
	methods for the modeling of reactive interfaces in photoelectrochemical
	({PEC}) solar cells,'' \emph{J. Comput. Phys.}, vol. 327, pp. 140--167, 2016.
	
	\bibitem{Ren2017}
	Q.~{Ren}, Y.~{Bian}, L.~{Kang}, P.~L. {Werner}, and D.~H. {Werner}, ``Leap-frog
	continuous–discontinuous {G}alerkin time domain method for
	nanoarchitectures with the drude model,'' \emph{J. Lightwave Technol.},
	vol.~35, no.~22, pp. 4888--4896, 2017.
	
	\bibitem{Chen2020APS_PML}
	L.~{Chen}, M.~B. {Ozakin}, and H.~{Bagci}, ``A low-storage pml implementation
	within a high-order discontinuous {G}alerkin time-domain method,'' in
	\emph{Proc. IEEE Int. Symp. Antennas Propag.}, 2020, pp. 1069--1070.
	
	\bibitem{Chen2020pml}
	L.~{Chen}, M.~B. {Ozakin}, S.~{Ahmed}, and H.~{Bagci}, ``A memory-efficient
	implementation of perfectly matched layer with smoothly-varying coefficients
	in discontinuous {G}alerkin time-domain method,'' \emph{IEEE Trans. Antennas
		Propag.}, pp. 1--1, 2020.
	
	\bibitem{Liu2003}
	T.-A. Liu, M.~Tani, and C.-L. Pan, ``{THz} radiation emission properties of
	multienergy arsenic-ion-implanted {GaAs} and semi-insulating {GaAs} based
	photoconductive antennas,'' \emph{J. Appl. Phys.}, vol.~93, no.~5, pp.
	2996--3001, 2003.
	
	\bibitem{Gobel2011}
	T.~{Gobel}, D.~{Schoenherr}, C.~{Sydlo}, M.~{Feiginov}, P.~{Meissner}, and
	H.~L. {Hartnagel}, ``Reliability investigation of photoconductive
	continuous-wave terahertz emitters,'' \emph{IEEE Trans. Microw. Theory
		Tech.}, vol.~59, no.~8, pp. 2001--2007, 2011.
	
	\bibitem{Ulbricht2011RMP}
	R.~Ulbricht, E.~Hendry, J.~Shan, T.~F. Heinz, and M.~Bonn, ``Carrier dynamics
	in semiconductors studied with time-resolved terahertz spectroscopy,''
	\emph{Rev. Mod. Phys.}, vol.~83, pp. 543--586, 2011.
	
\end{thebibliography}


\end{document}